\documentstyle[12pt,aaspp4,flushrt]{article}
\input psfig

\catcode`\@=11 
\def\lsim{\mathrel{\mathpalette\@versim<}}
\def\gsim{\mathrel{\mathpalette\@versim>}}
\def\@versim#1#2{\vcenter{\offinterlineskip
        \ialign{$\m@th#1\hfil##\hfil$\crcr#2\crcr\sim\crcr } }}

\newcommand{\rb}[1]{\raisebox{1.5ex}[0pt]{#1}}

\def\be{\begin{equation}}
\def\ee{\end{equation}}
\def\msun{M_\odot}
\def\sgr{Sgr A$^*$ }
\def\mdot{\dot m}
\def\ergs{\rm erg\,s^{-1}}
\def\sles{\lower2pt\hbox{$\buildrel {\scriptstyle <} 
   \over {\scriptstyle\sim}$}} \def\sgreat{\lower2pt\hbox{$\buildrel
{\scriptstyle >} \over {\scriptstyle\sim}$}}

\def\arcsec{^{\prime \prime}}
\def\arcmin{^{\prime}}
\def\degree{^{\circ}}

\begin{document}

\title{Advection-Dominated Accretion Model of Sagittarius A$^*$:
Evidence for a Black Hole at the Galactic Center}

\authoremail{authors@cfa.harvard.edu}
\author{Ramesh~Narayan, Rohan~Mahadevan, Jonathan~E.~Grindlay, 
Robert~G.~Popham, Charles~Gammie}

\affil{Harvard-Smithsonian Center for Astrophysics, 60 Garden St.,
Cambridge, MA 02138}

\begin{abstract}

Sgr A$^*$ at the Galactic Center is a puzzling source.  It has a mass
$M=(2.5\pm0.4)\times10^6M_\odot$ which makes it an excellent black
hole candidate.  Observations of stellar winds and other gas flows in
its vicinity suggest a mass accretion rate $\dot M\gsim {\rm
few}\times10^{-6}\msun\,{\rm yr^{-1}}$.  However, such an accretion
rate would imply a luminosity $> 10^{40}~{\rm erg\,s^{-1}}$ if the
radiative efficiency is the usual 10\%, whereas observations indicate
a bolometric luminosity $<10^{37}~{\rm erg\,s^{-1}}$.  The spectrum of
Sgr A$^*$ is unusual, with emission extending over many decades of
wavelength.  We present a model of Sgr A$^*$ which is based on a
two-temperature optically-thin advection-dominated accretion flow.
The model is consistent with the estimated $M$ and $\dot M$, and fits
the observed fluxes in the cm/mm and X-ray bands as well as upper
limits in the sub-mm and infrared bands; the fit is less good in the
radio below 86 GHz and in $\gamma$-rays above 100 MeV.  The very low
luminosity of Sgr A$^*$ is explained naturally in the model by means
of advection.  Most of the viscously dissipated energy is advected
into the central mass by the accreting gas, and therefore the
radiative efficiency is extremely low, $\sim5\times10^{-6}$.  A
critical element of the model is the presence of an event horizon at
the center which swallows the advected energy.  The success of the
model could thus be viewed as confirmation that Sgr A$^*$ is a black
hole.

\end{abstract}


\section{Introduction}

The enigmatic radio source, Sagittarius A$^*$ (Sgr A$^*$), has for
many years been a puzzle (Genzel \& Townes 1987; Genzel, Hollenbach \&
Townes 1994; Mezger, Duschl \& Zylka 1996).  The source is located at
the dynamical center of the Galaxy and is presumed to be associated
with a supermassive black hole.  However, observations provide
conflicting indications both for and against the black hole hypothesis
and there is currently no model that explains all the observations.
The paradoxical clues may be summarized briefly as follows:

\noindent
1. Dynamical measurements indicate a dark mass of $\sim(2.5\pm0.4)
\times10^6 M_\odot$ within the central 0.1 pc of the Galactic Center
(Haller et al. 1996; Eckart \& Genzel 1997).  This is believed to be
the mass $M$ of the putative supermassive black hole in Sgr A$^*$.

\noindent
2. Observations of stellar winds and gas flows near Sgr A$^*$, coupled
with the above estimate of the black hole mass, suggest that \sgr must
accrete gas from its surrounding at a mass accretion rate $\dot M
\gsim{\rm few}\times 10^{-6}M_\odot{\rm yr^{-1}}$ (Genzel et al. 1994).

\noindent
3. This $\dot M$ implies an accretion luminosity $L\sim0.1\dot Mc^2>
10^{40}~\ergs$, assuming a nominal radiative efficiency of 10\%.
However, \sgr is an unusually dim source with a total luminosity from
radio to $\gamma$-rays under $10^{37}~\ergs$.  The extremely low
luminosity has been used to argue against \sgr being an accreting
black hole (Goldwurm et al. 1994).

\noindent
4. For the above $M$ and $\dot M$ the peak emission from a standard
thin accretion disk will be in the near infrared (cf. Frank, King \&
Raine 1992).  In fact, this is a generic prediction of any model that
involves an optically thick flow radiating as a blackbody.  However,
Menten et al. (1997) have obtained a strong upper limit on the 2.2
micron flux of Sgr A$^*$ that effectively rules out such models.

\noindent
5. \sgr is brightest in the radio/mm band (see \S2 and Fig. 1), it is
weakly detected in X-rays (Predehl \& Trumper 1994), and it may have
been detected between 100 MeV and 2 GeV (Merck et al. 1996).  These
observations, combined with the Menten et al. infrared upper limit,
imply a spectral distribution completely unlike a blackbody or even a
sum of power-law components.  The observations suggest that \sgr is
optically thin, and that many different radiation processes may be in
operation.

Over the years a number of models have been proposed for Sgr A$^*$.
Some of these have been phenomenological approaches which aim to
explain the radio and infrared spectrum without including any detailed
dynamics (e.g. Falcke 1996, Duschl \& Lesch 1994, Beckert \& Duschl
1997).  These models generally require mass accretion rates lower than
those indicated by the observations.  Other models (Melia 1992, 1994;
Mastichiadis \& Ozernoy 1994) have attempted to incorporate dynamics,
but with simplifying assumptions such as ignoring the angular momentum
of the accreting gas.  One model of \sgr that has attempted a
self-consistent treatment of both viscous hydrodynamics and radiation
processes is the advection-dominated accretion flow (ADAF) model of
Narayan, Yi \& Mahadevan (1995).  An early and qualitative discussion
of this model was presented by Rees (1982).  The present paper is
based on the ADAF model.

An advection-dominated accretion flow is one in which most of the
energy released by viscous dissipation is stored in the gas and
advected to the center, and only a small fraction of the energy is
radiated (Narayan \& Yi 1994, 1995a, 1995b; Chen et al. 1995; see
Narayan 1997 for a recent review).  Most current work on ADAFs has
been concerned with a branch of low $\dot M$ solutions (Ichimaru 1977;
Rees et al. 1982; Narayan \& Yi 1995b; Abramowicz et al. 1995) which
is present for mass accretion rates below a few percent of the
Eddington rate (Narayan 1997; Esin, McClintock \& Narayan 1997).
These low-$\dot M$ solutions make use of the standard $\alpha$
viscosity and assume a {\it two-temperature} plasma with an
equipartition magnetic field.  The two-temperature paradigm was
introduced to astrophysics by Shapiro, Lightman \& Eardley (1976; see
also Phinney 1981 and Rees et al. 1982) and is based on the following
two assumptions: (1) It is assumed that the bulk of the viscous energy
is deposited in the ions, with only a small fraction of the energy
going directly into the electrons.  (2) Energy transfer from the ions
to the electrons is assumed to occur via Coulomb collisions, with no
significant nonthermal coupling being present (see Phinney 1981;
Begelman \& Chiueh 1988).  In the low-$\dot M$ two-temperature ADAF
model the ions achieve a nearly virial temperature,
$T_i\sim10^{12}~{\rm K}/r$, where $r$ is the radius in Schwarzschild
units, while the electron temperature saturates at around
$10^9-10^{10}$ K for $r\ \sles\ 100$ (Narayan \& Yi 1995b).  The high
ion temperature causes the gas to take up a quasi-spherical shape
(Narayan \& Yi 1995a), the gas is optically thin, and the flow does
not suffer from any serious thermal or viscous instabilities (Narayan
\& Yi 1995b; Abramowicz et al. 1995; Kato, Abramowicz \& Chen 1996).

The low-$\dot M$ branch of ADAFs has several appealing features for
explaining the observations of \sgr and other low-luminosity accretion
systems.  First, since most of the energy is advected with the gas and
lost into the black hole, the model naturally explains the low
luminosity of Sgr A$^*$.  Second, being optically thin, the spectrum
is quite different from blackbody.  Indeed, the high electron and ion
temperatures allow a variety of radiation processes to operate:
synchrotron, bremsstrahlung and inverse Compton from the electrons
(Narayan \& Yi 1995b; Mahadevan 1997), and $\gamma$-ray emission from
the ions via pion production (Mahadevan, Narayan \& Krolik 1997).
Therefore, the gas is likely to radiate over a wide range of
wavelengths.

The Narayan et al. (1995) ADAF model of \sgr gave a reasonable fit to
the observations available at that time and provided an explanation
for the low luminosity of the source.  However, the model required a
black hole mass of $7\times10^5M_\odot$ which differs from the most
recent dynamical mass estimate of $2.5\times10^6M_\odot$ (Eckart \&
Genzel 1997).  This problem prompted us to take another look at the
model.

In addition, there have been important developments on the
observational and theoretical front which further motivate the present
study.  The recent Menten et al. (1997) limit on the infrared flux of
\sgr is well below previous ``detections'' (Rosa et al. 1991; Eckart
et al. 1992) which were used by Narayan et al. (1995) to fit the ADAF
model.  Similarly, ASCA observations by Koyama et al. (1996) have
shown that the X-ray emission from the Galactic Center region is
dominated by an X-ray burster which lies within 1.3 arcminutes of
\sgr.  This implies that most previous measurements of the X-ray flux
of \sgr (Pavlinskii, Grebenev, \& Sunyaev 1992, 1994; Skinner et
al. 1987) are suspect since they were obtained with inadequate spatial
resolution.  Only the ROSAT PSPC detection by Predehl \& Trumper
(1994) survives as a reliable X-ray detection, although the
corresponding luminosity of the source in the ROSAT band is uncertain
due to uncertainties in the absorbing hydrogen column (cf. \S2).

The modeling techniques have also advanced significantly during the
intervening two years.  The Narayan et al. (1995) model was a local
one where the dynamics were calculated using a local self-similar
solution (Narayan \& Yi 1994) and the radiation was computed with a
local approximation for Compton scattering (Dermer, Liang \& Canfield
1991; Narayan \& Yi 1995b).  Consistent global dynamical solutions
have since been calculated, initially with a pseudo-Newtonian
potential (Narayan, Kato \& Honma 1997; Chen, Abramowicz \& Lasota
1997) and more recently for the full relativistic Kerr problem
(Abramowicz et al. 1996; Peitz \& Appl 1997; Gammie \& Popham 1997).
A global scheme for Comptonization has also been developed (Narayan,
Barret \& McClintock 1997), based on techniques due to Poutanen \&
Svensson (1996).  Finally, Nakamura et al. (1997) have shown that in
addition to energy advection by ions, which had been considered in
previous work, advection by electrons can be important in some
circumstances.  This effect is now included in the calculations
(Appendix A).  The present models are thus superior to those used by
Narayan et al. (1995).

In this paper we present improved ADAF models of Sgr A$^*$.  We make
use of the spectral data described in \S2 and employ the modeling
techniques outlined in \S3.  We present detailed results in \S4 and
conclude with a discussion in \S5.
 
\section{Spectral Data}

The radio to near infrared (NIR) spectrum of Sgr A$^*$ has been of
constant interest since the discovery of the source by Balick \& Brown
(1974), and observations have been carried out from 400 MHz (Davies et
al. 1976) to $\sim 10^{14}$ Hz (Menten et al. 1997).  Table 1 is a
compilation of the radio to NIR observations.  The data are also
plotted in Fig. 1.  The two general features in the table and figures
are 1) we have given only the maximum and minimum fluxes for
frequencies at which variability has been observed, and 2) open
circles correspond to low resolution measurements, which we treat as
upper limits, while filled circles correspond to the highest
resolution measurements in each band (subject to the additional
constraint that we require the resolution to be $<1$ arcmin).  The
latter measurements, which are identified with a $\star$ in Table 1,
are most likely to probe the actual accretion flow in Sgr A$^*$, while
the former could be contaminated by emission from other components in
the source.  This is discussed in detail below.

In determining the spectrum of the ADAF from Sgr A$^*$, special
attention has to be paid to the angular resolution of the
observations, which ideally must be comparable to the size of the
ADAF.  The angular size of the ADAF in \sgr depends on the wavelength
band under consideration.  The low frequency ($<1$ GHz) radio emission
is from radii on the order of $\sim10^3R_s-10^4R_s$ (where $R_s$ is
the Schwarzschild radius), which corresponds to an angular size at the
Galactic Center of $\lsim 0.06\arcsec$ ($\sim 500$ AU), where we have
taken $M = 2.5\times10^{6} M_{\odot}$.  With increasing frequency, the
size becomes smaller, and the sub-mm and infrared emission are from no
more than a few tens of $R_s$.  The soft X-rays are again from a large
volume (up to about $10^4R_s$), but harder X-rays are from
progressively smaller radii.  Gamma-ray emission from pion decay is
again limited to a few tens of $R_s$.

The radio spectrum of \sgr appears to consist of two components, with
a break at around the $\nu\sim86$ GHz point in Fig. 1.  The spectral
dependence is $L_{\nu}\sim \nu^{0.2}$ for $\nu< 40 $GHz, and steepens
to $L_{\nu}\sim \nu^{0.8}$ above 40 GHz. The steep increase has been
noted by Zylka et al. (1992, 1995), Serabyn et al. (1992, 1997) and
Rogers et al. (1995), but the origin of the two components is unclear.
Perhaps coincidentally, for frequencies $\lsim 50$ GHz, scattering by
turbulent density fluctuations in the ISM leads to source broadening.
The apparent size of the source, which is proportional to the square
of the observed wavelength (e.g. Davies et al. 1976; Backer 1982), is
larger than the intrinsic size at these frequencies.  The low
frequency data could thus in principle be contaminated by structures
in the source (e.g. jets) which are larger than the relevant radiating
region of the ADAF but smaller than the scatter-broadened size.

The ADAF model we present in the following sections fits the data
above 86 GHz well but underpredicts the flux at lower frequencies.
Motivated by this, and in view of the scattering argument mentioned
above, we plot the low frequency data as open circles in Fig. 1.  We
treat these data as upper bounds rather than as firm detections of the
central accretion flow in Sgr A$^*$.  Thus, even though the Marcaide
et al. (1992) and Alberdi et al. (1993) measurements at 22 GHz had
exceptionally good resolution, we still plot these points as open
circles because they are affected by scattering.  At 22 GHz, the
scattered source size is $\sim 15$AU, whereas the relevant region of
the ADAF which produces the radiation has a size of $\sim 1$AU.  Any
contamination to the radio flux must be present on a scale $\lsim 15$
AU, which is rather a stringent requirement.  We must therefore keep
open the possibility that the low frequency radiation (i.e. below the
apparent break at $\sim86$ GHz) also arises from the accretion flow.
The models presented in this paper are then unable to account for the
observed level of this low frequency emission.

For frequencies $\gsim 50$ GHz, the scattering size falls below the
intrinsic size.  High resolution observations can therefore determine
directly the emission from the central engine.  However, only a few
high resolution ($\lsim 1$ mas) observations have been made at these
wavelengths using VLBI or the VLBA.  These radio observations are
represented as filled circles in Fig. 1.  We take the point at 86 GHz
to be the best determined radio flux of \sgr for the purposes of
modeling the source.  Apart from the VLBI/VLBA points, all other radio
points $\gsim 86$ GHz have been represented by open circles since
these observations were done with poor resolution ($\gsim
4{\arcsec}$).

The FIR to NIR observations are all upper limits, and the low angular
resolution points are represented by open circles.  The upper limits
by Gezari et al. (1994) and Zylka et al. (1992) are the best
resolution observations in the FIR band, and are therefore shown as
filled circles.  Stolovy et al. (1996) detected a source in the
mid-infrared with an extinction-corrected flux of $100\pm40$ mJy.
However, their resolution was $0.5\arcsec$ and their flux may possibly
be contaminated by the compact cluster of sources observed by Eckart
et al. (1995).  We plot their data point as an upper limit with a
filled circle.  The recent upper limit of Menten et al. (1997) in the
NIR at 2.2$\mu$m is a very high resolution measurement since the
authors were able to determine accurately the relative position of
\sgr in their speckle images by comparing with a radio map of nearby
stellar SiO masers. The effective resolution of the speckle image is
$0.15\arcsec$, and therefore we plot this limit again as a filled
circle.

X--ray and hard X--ray as well as gamma-ray observations of Sgr A*
have long been limited by the relatively poor angular resolution
available at these energies and the very crowded field of the Galactic
Center region. Even at hard X--ray energies (e.g. 20--100 keV) where
the density of sources is relatively low, the Sgr A$^*$ source region
contains more sources ($\sim$10) in the central 5$\degree$ of the
Galaxy than any other region (Goldwurm et al. 1994).  At soft X--ray
(e.g. 0.5--4 keV) and low-medium X--ray energies (e.g. 2--10 keV),
this same region is even more crowded, with complex diffuse emission
and at least 10 sources within the central 1$\degree$ of Sgr A*
(cf. Watson et al 1991 and Koyama et al 1996). Thus the flux and
spectral distribution of the Galactic Center in the $\sim$1--100 keV
band, where the ADAF bremsstrahlung component peaks (cf. Fig. 1), is
particularly critical and angular resolution is paramount. At the soft
X--ray end of this band, the relatively uncertain absorbing column
density, $N_H$, towards the Sgr A$^*$ source is especially important
since the value chosen (which can in principle be determined by future
high resolution observations with AXAF) greatly affects the derived
source flux and luminosity.

We have included in the X--ray/hard X--ray region of the spectrum
plotted in Figure 1 only 4 points or upper limits which all are the
highest angular resolution available in their respective energy bands.
These data are listed in Table 2.  The one detection of a point-like
source with position fully consistent with Sgr A$^*$ is the ROSAT
detection (PSPC) with $\sim 20\arcsec$ resolution in the 0.8--2.5 keV
band reported by Predehl and Trumper (1994). This detection is plotted
for an assumed $N_H$ = 6 $\times$ 10$^{22}$ cm$^{-2}$, which is the
usual best estimate (cf. Watson et al 1981) for the interstellar
column density and which corresponds to the usually quoted visual
extinction of A$_V \sim$ 25--30 mag.  For this $N_H$, the total
integrated luminosity in the 0.8--2.5 keV band is $L_X = 1.5 \times
10^{34}$ erg s$^{-1}$.
\footnote{The $L_X$ and $\nu L_{\nu}$ calculations were done for the
ROSAT data using the PIMMS program, supplied by the HEASARC at GSFC.}
The $N_H$ used here differs from the much higher value (1.5 $\times$
10$^{23}$ cm$^{-2}$) assumed by Predehl and Trumper.  These authors
chose a higher $N_H$ in order to make the soft X--ray luminosity more
compatible with the variable compact source within $\sim 1\arcmin$ of
Sgr A$^*$ reported by Skinner et al (1987) and Pavlinskii, Grebenev
and Sunyaev (1992) from 2--20 keV coded aperture imaging observations
with modest (few arcmin) resolution.  However, recent ASCA
observations of the Galactic center in the 2--10 keV band, with $\sim
1\arcmin$ angular resolution but much higher spectral resolution, have
been reported by Koyama et al. (1996) and Maeda et al (1996). They
show the Sgr A$^*$ region to be complex, with diffuse emission over a
$\sim$2 $\times 3 \arcmin$ region and a point source at $1.3 \arcmin $
from Sgr A$^*$ which is an X--ray burster and therefore a neutron star
in a (8.4h) binary system. The burster has very likely dominated the
coded aperture imaging flux measurements. Koyama et al. therefore
quote an upper limit of 10$^{36}$ erg/s as the 2--10 keV luminosity
for the entire Sgr A$^*$ complex, with a possible actual value of
$\sim$10$^{35}$ erg/s for the point source alone. We have plotted this
ASCA upper limit in Fig. 1.

The ROSAT PSPC detection, although only with angular resolution of
$\sim 20\arcsec$, is a likely detection of Sgr A$^*$ with minimal
contamination from the surrounding diffuse source since it appears in
the ROSAT image to be consistent with a point source. However, higher
spatial resolution (e.g. AXAF) X--ray imaging, and/or evidence for
time variability, are needed to confirm this identification. In this
sense, even the ROSAT detection might be regarded as an upper limit.
Nevertheless, this data point is plotted as a solid symbol as it
corresponds to the highest resolution X-ray observation.  The
luminosity plotted in Fig. 1 and listed in Table 2 corresponds to the
likely $N_H$ value of $6\times10^{22}~{\rm cm^{-2}}$ and was obtained
by considering two extreme values of a power law spectral index:
photon index 1.0 and 2.0. These indices bracket the corresponding
index ($\sim$1.4) for a bremsstrahlung spectrum as predicted by our
model. The detected flux in the soft (0.8--2.5 keV) band is, however,
still highly sensitive to the assumed $N_H$ (as also pointed out by
Predehl and Trumper): for $N_H$ = 8 $\times$ 10$^{22}$ the flux
increases by a factor of 2.8 over that shown; whereas for $N_H$ = 5
$\times$ 10$^{22}$, it decreases by a factor of 0.6. Thus the fits to
ADAF models described in the next section, which are otherwise greatly
constrained by this ROSAT detection, should be regarded as uncertain
in normalization by a factor perhaps as large as 2 if uncertainties in
the interstellar $N_H$ are allowed for completely and perhaps as large
as 3 if internal self-absorption in the vicinity of the AGN is allowed
for (as suggested by Predehl and Trumper).  The box around the ROSAT
point in Fig. 1 includes both the uncertainty due to spectral index as
well as an additional factor of 2 uncertainty in normalizaton.  The
uncertain value of $N_H$ can, in principle, be directly measured or at
least greatly constrained by observations of Sgr A$^*$ with the ACIS
instrument on AXAF. This could observe the interstellar absorption
edge of Oxygen (at 0.8 keV) at the high angular resolution
(1$\arcsec$) needed to isolate completely the surrounding diffuse
emission.

At still higher energies, 20--100 keV, we plot the two upper limits
derived from the deep SIGMA observations of the Galactic Center
complex reported by Goldwurm et al. (1994). These observations are not
able to resolve the burster, and are likely affected by it (or,
rather, constrain the hard X--ray flux from the burster). More
sensitive observations in the hard X--ray band at energies above
60--100 keV, where the hard X--ray spectral component of neutron stars
in bursters usually is not detectable (e.g. Barret \& Grindlay 1995)
could provide a more stringent test of the ADAF spectrum as well as
the predicted spectral turnover.

The $\gamma$--ray spectrum shown in Fig. 1 corresponds to emission
from the EGRET source 2EG 1746--2852 which is coincident with the
Galactic Center (Merck et al. 1996).  The source is described as
point--like to within the resolution of the instrument ($ \sim
1\degree$), and is a significant excess ($\sim$10$\sigma$) above the
local diffuse emission.  The source has a very hard spectrum with a
photon index of 1.7 which differs significantly from other
unidentified Galactic EGRET sources.  The source might therefore
represent emission from Sgr A*, but it could equally well be
unresolved emission from a giant molecular cloud in the vicinity of
the Galactic Center.  If the source is Sgr A$^*$, the flux and
spectrum may be explained as $\gamma$-ray emission via pion production
and decay in the ADAF (Mahadevan et al. 1997).  If the source is
instead associated with a dense cloud of molecular hydrogen situated
at the Galactic Center, it will again show a pion decay spectrum but
with a spectral index (at energies above 100 MeV) given by the local
cosmic ray proton spectrum (i.e.  a power law with energy index
$\sim-1.6$). Due to the low resolution of EGRET ($1\degree$ beam) and
uncertainty in the nature of the source, we plot the $\gamma$-ray data
in Fig. 1 as open circles.  The EGRET observations (as reported by
Mayer-Hasselwander, 4th Compton Symp.)  indicate possible variability,
which would of course suggest that the source is indeed Sgr A$^*$, but
the same EGRET data also indicate that the source may be slightly
resolved, which would argue that the radiation is significantly
contaminated by surrounding diffuse emission.  Future observations
with higher sensitivity and better resolution are needed to resolve
this issue.

\section{Modeling Techniques}

We consider a black hole of mass $M$ accreting through an ADAF at a
rate $\dot{m} \dot{M}_{\rm Edd}$, where $\dot{M}_{\rm Edd} = 10 L_{\rm
Edd}/c^2 = 1.39 \times 10^{18}(M/\msun) ~{\rm g\,s^{-1}} = 2.21 \times
10^{-8}(M/\msun) ~\msun{\rm yr^{-1}}$.  An ADAF is in some sense
dynamically intermediate between a thin disk and a spherical accretion
flow: it is hot and quasi-spherical ($H/R \sim 1$, Narayan \& Yi
1995a), with rapid radial inflow, yet centrifugal support and viscous
angular momentum transport still play a significant role.  In
describing the ADAF we shall generally refer to the scaled radius $r
\equiv R/R_s$, where $R_s = 2 G M/c^2$ is the Schwarzschild radius.
We assume that the ADAF extends from the black hole horizon, $r=1$, to
an outer radius $r_{\rm out}$.  We take the mean angular momentum
vector of the ADAF to be inclined at an angle $i$ to the line of
sight.

To find the dynamical structure of the ADAF, we use the fully
relativistic, self-consistent, steady-state global solutions in the
Kerr metric developed by Popham \& Gammie (1997).  Their model uses a
nearly standard viscosity prescription which is parameterized by the
efficiency of angular momentum transport, $\alpha$ (assumed constant,
independent of radius).  The viscosity prescription has been modified,
however, so as to preserve causality and to include relativistic
effects.  We also use a quasi-spherical prescription for the vertical
structure based on Abramowicz, Lanza, and Percival (1997).  Comparison
of relativistic and nonrelativistic global models, and the
self-similar solution of Narayan \& Yi (1994) shows that they are
rather similar outside about $r = 5$.  Inside this radius the density
in our solution is very close to self-similar, but the temperature is
slightly lower.

The dynamical model is uniquely specified by four structure
parameters: $\alpha$, the viscosity parameter; $\gamma$, the adiabatic
index of the fluid, which is assumed to be a mixture of gas and
magnetic fields; $f$, the advection parameter, which gives the ratio
of advected energy to the viscous heat input; and $a$, the rotation
parameter of the black hole.  Of these four parameters, $\gamma$ is
determined in terms of the magnetic field equipartition parameter
$\beta$ (defined below) according to (Esin 1997): $\gamma=(8-3\beta)/
(6-3\beta)$.  The viscosity parameter $\alpha$ is again determined in
terms of $\beta$ following Hawley, Gammie \& Balbus (1996):
$\alpha=0.6(1-\beta)$.  The parameter $f$ is determined
self-consistently by iterating between the dynamical solution and the
radiation solution and seeking consistency between the two (see Esin
et al. 1997 for details).  The models considered here are extremely
advection-dominated, so that $f$ is very close to unity in all cases.
(For instance, for the baseline model shown in Fig. 1, we obtain
$f=0.9994$.)  Finally, we consider only the case a = 0, i.e. a
Schwarzschild black hole.

The thermodynamic state of the flow is described by the ion
temperature $T_i$, the electron temperature $T_e$, and the magnetic
pressure $p_{\rm mag} \equiv (1 - \beta) p_{\rm tot}$.  The total
pressure is fixed by the dynamical model.  We assume that $\beta$ is a
constant, independent of radius, and so $\alpha$ and $\gamma$ are also
independent of radius.  Thus, all the structural parameters are
constant in our model (cf. \S3).  Given $\beta$ we know the run of gas
pressure with radius.  Since the electrons can cool efficiently and
coupling between the ions and electrons is assumed to be weak
(provided solely by Coulomb collisions), we have $T_i \gg T_e$.  The
run of gas pressure then gives $T_i$ to a good approximation.

The heart of the problem is now calculating the electron temperature,
which determines the spectral properties of the flow.  A detailed
description of our procedure is given in Esin et al. (1997).
Basically, the electron temperature is determined by solving an energy
balance equation:
\be
Q^{\rm e,adv} = Q^{\rm ie} + \delta\,Q^{\rm vis} - Q^{\rm rad}.  
\ee 
Here $Q^{\rm e,adv}$ is the rate at which energy in the electrons is
advected inward by the flow.  This term is discussed in greater detail
in Appendix A.  Our earlier work (with the exception of Esin et
al. 1997) ignored the advective term, but recently Nakamura et
al. (1997) (see also Mahadevan \& Quataert 1997) have shown that it
can be important, particularly at low mass accretion rates.  At low
$\mdot$ the electrons are nearly adiabatic and the compressive heating
as the gas flows in becomes large.  The other terms in equation (1)
are as follows: $Q^{\rm ie}$ is the rate at which the electrons are
heated by Coulomb collisions, $Q^{\rm vis}$ is the total rate of
viscous dissipation, and $Q^{\rm rad}$ is the radiative cooling.  The
parameter $\delta$ describes the fraction of the viscous heating that
goes into the electrons.  In general we set $\delta = 10^{-3} \sim
m_e/m_p$.

Because of Compton scattering, the radiative cooling term couples the
flow at different radii.  It is calculated using the iterative
scattering method described in detail in Narayan, Barret \& McClintock
(1997).  The ADAF is represented by a logarithmically spaced grid of
nested shells.  Within each shell all flow variables are assumed
uniform for $\pi/2 - \theta_H < \theta < \pi/2 + \theta_H$, where
$\theta_H$ is an effective angular scale height, calculated as in
Narayan, Barret \& McClintock (1997) Appendix A.  We then guess a
value of the electron temperature and compute the synchrotron and
bremsstrahlung emission from each shell.  Then, using the probability
matrix elements $P_{ij}$, which give the probability that a photon
emitted in shell $i$ of the ADAF is scattered by an electron in shell
$j$, the rate of cooling through Compton scattering is calculated (see
Narayan, Barret \& McClintock 1997 for details).  This procedure is
iterated until convergence is achieved.  The iteration method ensures
that multiple scattering within the ADAF is properly taken into
account.  It also ensures that the advection parameter $f$ used in the
dynamical solution is consistent with the actual energy advection, as
determined by the radiation solution.  Doppler shifts and ray
deflections are ignored in the present calculations.

Once the cooling term is calculated, we must get the photons out to
the observer.  We use essentially Newtonian photon transport, although
we have taken a step toward relativistic photon transport by including
gravitational redshift (this is also included in the Compton
scattering calculation).  We do not include Doppler shifts, which to
lowest order in $v/c$ broaden the spectrum.  Since all features in our
spectra have $\delta\nu/\nu$ of order unity, this should not greatly
alter the gross properties of the spectrum, except where it is very
steep.  At next higher order in $v/c$ gravitational redshift, higher
order Doppler, and various geometric effects all enter.  In our model
we have included a correction for the volume of the emitting region
($1/\sqrt{1 - 1/r} \times$ the Euclidean value) and for gravitational
redshift.  The sum of these effects is that $L_\nu(\nu)$ observed at
large radius is now $L_\nu(\nu \sqrt{1 - 1/r})$, that is, $L_\nu$ is
shifted redward by the gravitational redshift factor.  Jaroszynski \&
Kurpiewski (1997) have recently considered the full effects of photon
transport near a Kerr black hole on the spectrum of an ADAF.

The above discussion is concerned with radiation from electrons.  We
also compute pion production by the hot protons in the ADAF and the
resulting $\gamma$-ray emission through pion decay.  The procedure we
follow for this calculation is described in Mahadevan et al. (1997)
and is based on earlier work by Dermer (1986).  The present
calculations differ in three respects from those in Mahadevan et al.:
1) the density profile of the protons is now given by the relativistic
global solution of Popham \& Gammie (1997) instead of the
pseudo-Newtonian solution of Narayan, Kato \& Honma (1997), 2) we
allow for the non--spherical geometry of the flow via the angle
$\theta_H$ mentioned above, and 3) we include the effect of
gravitational redshift.

The pion production rate depends sensitively on the energy
distribution of the protons.  Mahadevan et al. (1997) considered two
extreme distributions: a thermal distribution and a power-law
distribution.  The latter has significantly more pion production and
maximizes the efficiency of $\gamma$-ray emission.  We employ the
power-law model with an energy index $p=2.3$.  The choice of model is
dictated by the EGRET spectrum of the Galactic Center, which is
consistent with a power--law distribution of proton energies and is
inconsistent with a thermal distribution (Mahadevan et al. 1997).
Note that Mahadevan \& Quataert (1997) have shown that the protons in
an ADAF do not have time to thermalize; therefore, it is permissible
to assume a nonthermal energy distribution for the protons.

\section{Results}

In this section we combine the modeling techniques outlined in \S3
with the observational constraints described in \S2 to come up with
ADAF models of Sgr A$^*$.  Our standard parameters are as follows.  We
assume that the ADAF extends from $r=1$ to $r_{\rm out}=10^{5}$.  For
the black hole mass, we take the estimate of Eckart \& Genzel (1997):
$M=2.5\times10^6M_\odot$.  We assume exact equipartition between gas
and magnetic pressure in the accreting gas: $\beta=0.5$.  Following
Hawley, Gammie \& Balbus (1996) we take the viscosity parameter to be
$\alpha\sim0.6(1-\beta)=0.3$ (the coefficient lies in the range
$0.5-0.6$).  We assume that electrons receive only a small fraction
$\delta\sim m_e/m_p$ of the viscous dissipation as direct heating:
$\delta=0.001$.  We assume a Schwarzschild black hole: $a=0$.  Since
we do not know the orientation of the angular momentum vector of the
accreting gas, we set the inclination of the system to a generic
value: $i=60\degree$.  The models considered here are extremely
optically thin and so the results are virtually independent of $i$.

The only parameter that needs discussion is $r_{\rm out}$.  In a
viscous rotating accretion flow, the outer edge of the flow is
determined by the radius at which the outward viscous transport of
angular momentum is balanced by external torques.  In the case of \sgr
it is not clear exactly where this balance is achieved, but it is
likely to be roughly at the radius where the mass supply originates.
Our choice of $r_{\rm out}=10^5$, corresponding to $R_{\rm out}=0.024$
pc, may be an underestimate of the true outer radius.  However, this
is not a serious concern for the modeling since for such large outer
radii the results are insensitive to the actual value chosen.  This
is because the energetically important region of the flow is closer to
the black hole.

As described above, all the parameters of the model are fixed at
standard values in our model.  The only parameter that we consider
fully adjustable is the mass accretion rate $\mdot$.  In each model
described below, we have adjusted $\mdot$ so as to fit the ROSAT X-ray
flux (but see the discussion in \S2 for uncertainties in the X-ray
flux due to the hydrogen column).  Although we treat $\mdot$ as a free
parameter, there are in fact some constraints.  The infrared source
IRS 16 appears to be the primary supplier of gas to Sgr A$^*$.
Assuming that \sgr captures a fraction of the wind of IRS 16 by Bondi
accretion, Melia (1992) estimated $\dot M\approx
2\times10^{-4}~M_\odot{\rm yr^{-1}}$ assuming a wind velocity of
$600~{\rm km\,s^{-1}}$, while Genzel et al. (1994) estimated a mass
accretion rate of $\dot M\approx 6\times10^{-6}~M_\odot{\rm yr^{-1}}$
using a wind velocity of $1000~{\rm km\,s^{-1}}$ .  We take these two
estimates to be low and high extremes.  Converting to Eddington units,
$\mdot$ is thus likely to be in the range
$10^{-4}<\mdot<3\times10^{-3}$.

The solid line in Fig. 1 shows our baseline model where $M$, $\alpha$,
$\beta$, and $\delta$ have been set to their standard values
($2.5\times10^6M_{\odot}, ~0.3, ~0.5, ~0.001$ respectively) and
$\mdot$ has been adjusted to fit the X-ray flux.  This model has a
mass accretion rate of $\mdot=1.3\times10^{-4}$, which lies within the
acceptable range of $\mdot$ discussed above.  The computed spectrum
has four well-defined peaks.  From the left, these correspond to
self-absorbed thermal synchrotron emission, singly Compton-scattered
synchrotron radiation, bremsstrahlung emission, and $\gamma$-rays from
neutral pion decay.  The model spectrum passes through the VLBI radio
flux measurement at 86 GHz, which we have identified to be a reliable
high frequency radio observation (cf. \S2), and turns over sharply in
the sub-millimeter band exactly as required by the sub-millimeter and
infrared upper limits.  The model also satisfies the ROSAT X-ray
detection and the other X-ray upper limits.  Considering that only one
parameter, $\mdot$, has been adjusted, we consider the agreement with
the data satisfactory.

Note that the infrared and X-ray fluxes have come down significantly
compared to the data shown in Narayan et al. (1995).  The present
model is compatible with both the new measurements.  Interestingly, we
cannot find any model that fits either the new infrared upper limit
with the old X-ray data, or the new X-ray flux with the old infrared
data.

The model does not agree with the low frequency radio measurements at
frequencies below 86 GHz.  We discuss this discrepancy at the end of
this section.

The model also has a discrepancy in the $\gamma$-ray band; the
predicted flux is lower than the observations by approximately an
order of magnitude.  The pion decay model requires an accretion rate
of $\mdot=4.5\times10^{-4}$ to fit the observed flux whereas the fit
to the rest of the spectrum gives $\mdot=1.3\times10^{-4}$.  There is
thus a discrepancy of a factor $\sim3-4$ between the two values of
$\mdot$.  We have been unable to come up with a reasonable resolution
of this discrepancy.  Considering that we have used an extreme model
which maximizes the pion production rate (cf. \S2), we are compelled
to suggest that perhaps the EGRET source does not correspond to Sgr
A$^*$, but rather to a concentration of molecular gas at the Galactic
Center.  The resulting emission is then expected to be much more
diffuse, but it would still appear unresolved to the $1\degree$ beam
of EGRET.

Two features of the ADAF model in Fig. 1 should be highlighted.
First, the model fits the data using a reasonable mass for the black
hole, $M=2.5\times10^6M_\odot$.  This is an improvement over the model
described in Narayan et al. (1995) where the data could be fitted only
with a mass $M=7\times10^5M_\odot$.  The primary reason for the
improvement is the inclusion of electron energy advection, or
compressive heating of the electrons (Nakamura et al. 1997), as we
explain below.  Second, the model is extremely advection-dominated.
The bolometric luminosity $L_{\rm bol}$ integrated over all
frequencies is only $2.1\times10^{36}~\ergs$, which corresponds to a
radiative efficiency, $\epsilon=L_{\rm bol}/\dot Mc^2=5\times10^{-6}$.
It is this extraordinarily low radiative efficiency that allows the
model to fit the observations with such a large mass accretion rate.

In contrast, a standard thin accretion disk model of \sgr is ruled out
quite comprehensively by the radio--IR--X-ray data.  The dotted line
in Fig. 1 shows a thin disk model with $\mdot=10^{-4}$, the lowest
mass accretion rate that we consider reasonable.  The spectrum was
calculated assuming that the emission is blackbody at each radius.
The model predicts an infrared flux which is many orders of magnitude
above the measured upper limits.  The short dashed line in Fig. 1
shows another thin disk model where $\mdot$ has been reduced to
$10^{-9}$.  This model does satisfy the IR upper limits but does not
fit any of the observations, either in the radio or X-ray bands.
Further, the mass accretion rate is unreasonably small.

A number of improvements have taken place in the modeling techniques
since the publication of our first model of \sgr (Narayan et
al. 1995).  It is interesting to investigate what effect each
improvement has had on the calculated spectrum.  Figure 2 shows a
sequence of models in which we start with the simplest version of the
model and progressively add features one by one.  Figures 3--5 explore
the effects of varying model parameters and are described below.

The dotted line in Fig. 2a corresponds to the most primitive version
of the ADAF model, in which the flow is assumed to have a self-similar
form (Narayan \& Yi 1994) and neither compressive heating of electrons
nor gravitational redshift is included.  The optimized accretion rate
is $\mdot=6\times10^{-5}$.  The general shape of the spectrum is
similar to that of our standard model (the solid line), but this model
differs in three ways.  First, the emission is stronger in nearly all
bands compared to our standard model; in fact, the model is
inconsistent with the infrared limit.  Second, the mass accretion rate
is lower than in our standard model by a factor of 2.  This, combined
with the higher luminosity, means that the model is not as
advection-dominated as our standard model; we obtain a radiative
efficiency of $\epsilon=2.4\times10^{-4}$.  Third, the calculated
synchrotron emission is well above the 86 GHz VLBI point.  The
discrepancy in the radio flux is more than an order of magnitude and
is serious.  Within the framework of the self-similar flow assumption,
the only way to eliminate this problem is by changing one or more of
the model parameters.  This was, in fact, the primary reason why the
Narayan et al. (1995) model required a black hole mass of
$7\times10^5M_\odot$ instead of the current best estimate of
$2.5\times10^6M_\odot$ (see Fig. 3 below which shows how the spectrum
is modified when the black hole mass is reduced).

The short-dashed line in Fig. 2a shows the effect of replacing the
self-similar flow by a global flow based on a pseudo-Newtonian
potential (Narayan, Kato \& Honma 1997; Chen, Abramowicz \& Lasota
1997).  The changes are minimal.

The long-dashed line in Fig. 2a next shows the effect of including
compressive heating (Appendix A and Nakamura et al. 1997).  This
model, which has $\mdot=1.1\times10^{-4}$, has a very different
spectrum compared to the previous two models.  The overall emission is
significantly reduced, notably in the synchrotron peak, and the flow
is substantially more advection-dominated:
$\epsilon=2.1\times10^{-6}$.  This model fits the 86 GHz data point
very well without requiring any adjustment to the black hole mass.  We
thus confirm the result of Nakamura et al. (1997) that compressive
heating is an important effect.  Figure 2b shows the electron
temperature profiles of the various models.  The model with
compressive heating has a significantly different temperature
structure than the models without.  At very low $\mdot$, as in our
model of Sgr A$^*$, the dominant terms in the electron energy equation
are the two pieces of $Q^{\rm e,adv}$ written in the Appendix,
viz. the terms proportional to $dT_e/dR$ and $d\rho/dR$.  These two
dominant terms balance each other, while the rest of the terms in
equation (1) are small.  In other words, the electrons are essentially
adiabatic.  The adiabatic condition gives a lower electron temperature
than in the previous two models and this accounts for the change in
the shape of the synchrotron contribution to the spectrum.

The dash-dot line in Fig. 2a shows next the effect of including a
fully relativistic global solution ($\mdot=1.3\times10^{-4}$), taken
from Popham \& Gammie (1997), instead of the pseudo-Newtonian global
solution employed in the previous two models.  Most of the features
are similar, but the overall emission is increased because the
relativistic solution has a lower radial velocity close to the black
hole compared to the pseudo-Newtonian solution.  (This is because
$v<c$ at all radii in the relativistic model whereas the
pseudo-Newtonian model gives $v>c$ close to the black hole.)
Consequently, the density is higher in the relativistic model and this
leads to increased emission.

Finally, the solid line in Fig. 2a shows our standard model, which
includes gravitational redshift.  As expected, this model has a lower
luminosity than the previous model, but is otherwise quite similar.

Figures 3--5 show the effect of varying the other parameters in the
model.  Figure 3a shows what happens when we vary the mass.  In
addition to the baseline model, we present two other models with $M =
1/3$ and $3$ times $M$ of the baseline model.  As Mahadevan (1997) has
shown, with increasing $M$, the magnetic field strength decreases as
$M^{-1/2}$ and this causes the synchrotron peak to move to lower
frequencies $\propto M^{-1/2}$.  Of the three models shown in Fig. 3a,
the one with $M=2.5\times10^6M_\odot$ gives the best fit to the VLBI
radio data.  Assuming our choices of $\alpha$ and $\beta$ are correct,
this provides additional support for the black hole mass measured by
Haller et al. (1996) and Eckart \& Genzel (1997).

Figure 3b shows the effect of varying the mass accretion rate.  In
addition to the baseline model, four other models are shown, with
$\mdot=1/2, ~1/\sqrt{2}, ~\sqrt{2}, $ and $~2$ times $\mdot$ of the
baseline model.  The radiative efficiency of ADAFs varies rapidly with
$\mdot$: $\epsilon\propto\mdot$ (Narayan \& Yi 1995b).  Therefore, the
luminosity varies as $\mdot^2$.  This can be seen in both the
synchrotron and bremsstrahlung peaks.  The Compton peak in the
infrared shows an even stronger dependence on $\mdot$.  This is
because the Comptonized flux is proportional to the product of the
synchrotron emission (which is $\propto \mdot^2$) and the optical
depth (which is $\propto\mdot$), so that the amplitude of this peak
varies approximately as $\mdot^3$.  The three intermediate models in
Fig. 3b are consistent with all the data, but the two extreme models
lie outside the X-ray error box and one of them also violates the
sub-millimeter and infrared limits.  Since the X-ray luminosity of
\sgr is uncertain, Fig. 3b indicates how the model $\mdot$ will need
to be modified if a future determination of $N_H$ leads to a
significant revision of the X-ray flux.

Figure 4a shows the effect of varying $\alpha$.  For each $\alpha$, we
have adjusted the accretion rate to fit the X-ray flux.  The spectrum
is not very sensitive to $\alpha$ (except in the optical/IR), though
the fitted mass accretion rates show modest variations:
$\mdot=6.3\times10^{-4}, ~1.0\times10^{-3}, ~1.3\times10^{-3},
~1.5\times10^{-3}$ for $\alpha=0.1,~0.2, ~0.3,~0.4$ respectively.

Figure 4b shows the effect of varying $\beta$.  Again, for each
$\beta$, we have adjusted the accretion rate to fit the X-ray flux.
In this case, we see quite substantial changes in the predicted
spectrum.  An increase in $\beta$ leads to a decrease in the magnetic
field strength and one may be tempted to think that this would cause
the synchrotron peak to reduce in amplitude.  In fact, the opposite
behavior is seen.  The reason can be traced to compressive heating.
Since the electrons are effectively adiabatic, their temperature
profile is determined by their adiabatic index $\gamma_e$.  We show in
Appendix A that because the gas is a mixture of particles and magnetic
field, the effective $\gamma_e$ depends on the parameter $\beta$.  As
$\beta$ goes up, the magnetic pressure goes down and $\gamma_e$
increases.  A larger $\gamma_e$ leads to hotter electrons, see
Fig. 4c, which causes more synchrotron emission.  Unfortunately, this
means that the results are sensitive to the details of how we model
energy advection in the electrons.  In this context we note that our
equation of state for the electrons differs somewhat from the one used
by Nakamura et al. (1997).  We use a relativistic equation of state
which causes $\gamma_e$ to change as the electron temperature
approaches and crosses the value $T_e\sim m_ec^2/k$.  In addition, we
modify the adiabatic index of the electrons to allow for the
equipartition magnetic field which is assumed to be present in the gas
and is coupled to the electrons.  Nakamura et al. (1997) treat the
particles as a separate component and ignore the field in their
equation of state.

Figure 4d shows the effect of varying $\delta$.  For all $\delta\
\sles\ 0.01$ the spectrum is generally unaffected.  This result is
different from that found in Narayan, Barret \& McClintock (1997).
Once again, the reason is the inclusion of compressive heating.
Because the electrons are now essentially adiabatic, their temperature
profile is not affected by modest changes in the heating or cooling.
Only when $\delta$ is large, e.g. $\delta=0.03162$ (the dot-dashed
line in Fig. 4d), do the electrons experience significant additional
heating and only then does the spectrum show a noticeable change.

We turn now to the low frequency radio data, where the model deviates
substantially from the measurements.  As discussed in \S2, this region
of the spectrum may possibly arise from a separate component such as a
jet which is outside the scope of our model.  The problem with the jet
proposal is that the additional component has to be quite compact
($\lsim15$ AU) in order not to be seen in the high resolution VLBI
images of Marcaide et al. (1992) and Alberdi et al. (1993) at 22 GHz.
If the low frequency radio emission is from the accretion flow itself,
then we need to identify a mechanism which could increase the radio
emission over and above our baseline model.

In our model, the synchrotron radiation at different frequencies are
produced at different radii in the flow, the emission at $10^{12}$ Hz
coming from close to the black hole and the emission at lower
frequencies coming from farther out.  Thus, one simple way of
improving the fit to the radio data is to modify the electron
temperature outside of a few tens of Schwarzschild radii.  Figure 5 is
{\em only} an illustration.  The solid line in Fig. 5a corresponds to
our baseline model, while the dashed line is another model which is
identical in all respects except that we have arbitrarily set
$T_e=2\times10^9$ K over the radius range $r=20-1000$ (Fig. 5b).  This
ad hoc model fits the data well.

Is there any reason to think that the electrons might have the profile
shown in Fig. 5b?  The answer is tentatively yes, since there are
several radial transport mechanisms which could drive the electrons to
a nearly isothermal state.  First, the long mean free path of the
electrons can lead to fairly strong radial heat conduction (parallel
to field lines).  Second, synchrotron radiation, which has a
thermalizing effect on the electrons (Mahadevan \& Quataert 1997), can
also cause significant energy diffusion.  Finally, the tangled
magnetic fields which we assume to be present in the flow may move
outwards as a result of buoyancy and may dissipate their energy and
heat the electrons at larger radii.  Some of these effects could well
be episodic; it is interesting in this connection that the radio flux
of \sgr is known to be quite variable (Zhao et al. 1989).  These
effects need further study.

\section{Discussion}

The main result of this paper is that the two-temperature ADAF model
(Narayan et al. 1995, Rees 1982) provides a viable explanation of the
spectral properties of Sgr A$^*$.  Our basic model, shown by the solid
line in Fig. 1, fits most of the high resolution measurements from the
high frequency ($\geq86$ GHz) radio to the X-ray band, including the
stringent infrared limit of Menten et al. (1997) and the revised X-ray
luminosity.

The ADAF model described in this paper makes use of a two-temperature
plasma, as originally described by Shapiro et al. (1976).  This
involves two key assumptions.  First, it is assumed that most of the
energy released by viscosity goes into the ions and not the electrons.
Second, the only coupling between ions and electrons is assumed to be
via Coulomb collisions.  At the present time, these assumptions have
not been proved theoretically to be valid, nor have they been
disproved (e.g.  Begelman \& Chuieh 1988).  Rees et al. (1982) made
the interesting point that the plasma physics involved is so complex
that perhaps the best way of testing the two-temperature paradigm is
by comparison of astrophysical models with observations.  Fabian \&
Rees (1995) declared, based on the success of our earlier model of
\sgr (Narayan et al. 1995), that the two-temperature assumption
appears to be supported by observations.  As the present paper shows,
the two-temperature ADAF model explains the data on \sgr even better
than it seemed in the 1995 paper.  Furthermore, the ADAF model has
been successfully applied to several other sources (see Narayan 1997
for a review), making the Fabian \& Rees argument even stronger.

Although the ADAF model presented here includes a number of
parameters, only {\it one} parameter is adjusted in the spectral fits,
and the quality of the agreement between the model and the data in
Fig. 1 must be judged in this context.  The free parameter in the
model is the Eddington-scaled mass accretion rate $\mdot$ which is
adjusted so as to fit the X-ray luminosity in the ROSAT band.
Although there is some uncertainty in the X-ray luminosity because of
uncertainty in the hydrogen column (cf. \S2), we feel this does not
seriously affect the model fit (see Fig. 4c which shows how small
adjustments in the fitted parameter $\mdot$ will allow the model to
remain consistent with any future revisions of the X-ray luminosity).

Apart from $\mdot$ all the other parameters in the model are assigned
standard values (these values are not changed from one application to
another).  For instance, we invoke strict equipartition and assign
$\beta=0.5$.  Once $\beta$ is given, the model has a unique
prescription (Esin 1997) for the adiabatic index of the gas: $\gamma=
(8-3\beta)/(6-3\beta)$.  Further, by making use of the scaling between
the shear stress and the magnetic pressure derived by Hawley, Gammie
\& Balbus (1996), we are also able to fix the viscosity parameter:
$\alpha\sim0.6(1-\beta)$.  Of course, the use of the $\alpha$
prescription itself involves an assumption about the nature of the
viscosity.  Narayan (1996), however, has argued that this viscosity
prescription with $\alpha$ independent of radius is particularly
appropriate for ADAFs.  The final parameter in the model is $\delta$,
the fraction of the viscous energy that goes directly into the
electrons.  By assumption $\delta$ is small, and it is usually
assigned a value $\delta=10^{-3}\sim m_e/m_p$.  The results are,
however, insensitive to the exact value so long as
$\delta\lsim10^{-2}$ (see Fig. 4d).

The models presented here are substantially more elaborate (cf. \S2)
than the ones described in Narayan et al. (1995).  It is pleasing that
the more detailed analysis presented here leads to an improvement in
the agreement with observational constraints.  In contrast to the
models presented in Narayan et al. (1995), where a black hole mass of
$7\times 10^5M_\odot$ was required in order to accommodate the radio
data, here we find that the model quite naturally fits the data with a
more reasonable mass of $2.5\times10^6M_\odot$ (Eckart \& Genzel
1997).  This removes one of the main problems with the previous work.
The model also predicts a mass accretion rate $\mdot\sim
1.3\times10^{-4}$ (in Eddington units) which is within the range
considered likely by a direct estimate of the accretion rate (Genzel
et al. 1994).

We should emphasize that the model is self-consistent, and includes a
detailed treatment of hydrodynamics, radiation processes, and thermal
balance of ions and electrons in the two-temperature plasma.  Most
other models in the literature are more primitive and/or ad hoc.  In
addition, the model is fairly robust to changes in the parameters
(Fig. 4).  The predicted spectrum is insensitive to large variations
in the viscosity parameter $\alpha$ and the electron heating parameter
$\delta$.  The results do, however, vary with the equipartition
parameter $\beta$ (Fig. 4b, 4c).

There are some problems with the present model.  First, we
under--predict the $\gamma$-ray flux in the EGRET band by an order of
magnitude.  This might indicate a residual uncertainty in the relative
normalization of the protons and electrons, but we have not been able
to come up with any specific proposal to eliminate the discrepancy.
It is worrying that we underpredict the flux even though we have
assumed a power-law energy distribution for the protons which
maximizes the predicted $\gamma$-ray luminosity.  Perhaps the
$\gamma$-ray source detected by EGRET with its one degree beam is not
Sgr A$^*$ at all, but diffuse emission from a molecular cloud in the
Galactic Center region.

Another problem is that the model under--predicts the radio flux below
86 GHz.  The existence of a break in the radio spectrum at around 86
GHz suggests that the region of the spectrum below 86 GHz might arise
from a distinct component, e.g. a weak outflow of some kind, as
postulated in the model due to Falcke (1996).  We note in this
connection that Krichbaum et al. (1993) claim to have seen evidence
for a jet in their 43 GHz VLBI image, though Backer et al. 1993 could
not verify this.  One problem with the jet hypothesis is that the 22
GHz VLBI maps of Marcaide et al. (1992) and Alberdi et al. (1993)
show no evidence for any resolved structure.  This means that the jet
has to be smaller than 15 AU, i.e.  no larger than ten times the size
we would predict for the ADAF at this frequency.  If the radio flux is
from a jet, then clearly the discrepancy we face in the radio
luminosity is not an issue since our model includes only the accretion
flow.  However, if the radiation is not from a separate component but
from the ADAF itself, then it probably implies that the electron
temperature in the flow differs from that predicted by our model.
Figure 5 shows the kind of temperature profile that is needed to fit
the observations.  The profile shown here is completely ad hoc.
However, if radial transport processes are included, the model might
give a qualitatively similar temperature profile.  We would emphasize
that the luminosity of \sgr in low frequency radio waves is a small
fraction (less than a percent) of the bolometric luminosity of the
source.  Minor changes in the flow profile can explain the discrepancy
here without affecting the overall flow dynamics or energetics.

Yet another problem is the somewhat low electron brightness
temperatures predicted by the model.  As Fig. 4c shows, we obtain
values below $10^{10}$ K at all radii, whereas VLBI observations at 43
GHz and 86 GHz indicate brightness temperatures in excess of $10^{10}$
K (Backer et al. 1993, Rogers et al. 1994).  Thus, although the model
fits the 86 GHz flux well, it seems to predict a larger angular size
for the source, by a factor of $\sim 1.5$, than observations indicate.
For comparison with future observations, we list here linear source
radii at various radio frequencies according to the baseline model
shown in Fig. 1: $1.3\times10^{12}$ cm at $10^{12.5}$ Hz,
$4.1\times10^{12}$ cm at $10^{12}$ Hz, $1.0\times10^{13}$ cm at
$10^{11.5}$ Hz, $2.1\times10^{13}$ cm at $10^{11}$ Hz,
$4.1\times10^{13}$ cm at $10^{10.5}$ Hz, $8.3\times10^{13}$ cm at
$10^{10}$ Hz.  Note that the electron temperature depends on the value
of the parameter $\beta$ (Fig. 4c).  Thus, one might be able to fit
the observed brightness temperatures by tuning this parameter as a
function of radius, although we have kept $\beta$ fixed in our models.

An outstanding feature of the ADAF model presented here is its
extraordinarily low radiative efficiency: $\epsilon=L_{\rm bol}/ \dot
Mc^2=5\times10^{-6}$.  The low efficiency permits the model to fit the
very low luminosity of \sgr with a fairly large $\mdot$.  Figure 1
shows two thin disk models with the standard radiative efficiency of
$\epsilon\sim0.1$.  A thin disk with the ``correct'' $\mdot$ (dotted
line) over--predicts the infrared flux by four or five orders of
magnitude.  A model with $\mdot$ reduced by a factor of $10^5$
accommodates the IR limits but does not fit any of the other data and
has a mass accretion rate which is extremely discrepant with
independent estimates of $\mdot$.  Some models in the literature which
are based on a thin disk attempt to solve the luminosity problem by
hiding a large fraction of the emission in the optical band where
there is extinction by dust (e.g. Falcke et al. 1993a, Melia 1994).
However, the models invariably predict emission in the infrared at a
level well above the Menten et al. (1997) limit.  Other models in the
literature which are not based on a thin disk also have problems
fitting the low luminosity (e.g. Duschl \& Lesch 1994; Mastichiadis \&
Ozernoy 1994) and generally require values of $\mdot$ significantly
below the lower limit of $10^{-4}$.

Can we save the thin disk model by assuming that the disk has no
viscosity at all, so that the gas in the disk does not accrete at the
present time?  In a very interesting argument, Falcke \& Melia (1997)
show that even such an extreme model is inconsistent.  This is because
the mere presence of the disk, even though it does not accrete, will
still lead to fairly strong infrared emission.  The argument is that
if there is inflow of material towards the Galactic Center via a
Bondi-Hoyle-like flow, when the inflowing gas hits the disk and
circularizes at a radius $r_{\rm circ}$ it will produce a substantial
amount of thermal radiation.  For the mass accretion rate of $\mdot\
\sgreat\ 10^{-4}$ estimated in the case of \sgr (cf. \S4) the
predicted infrared flux is well above the Menten et al. (1997) limit
in nearly all the cases considered by Falcke \& Melia (1997).  The
only way of avoiding this argument is by making $r_{\rm circ}$ large
(say $r_{\rm circ}\gsim10^4$), but Falcke \& Melia show that this is
not consistent with most reasonable wind specific angular momentum
configurations.

The ADAF model is able to circumvent the Falcke \& Melia argument
quite naturally and this is another point in favor of this model.
Since the accreting gas in the ADAF has a quasi-spherical shape
(Narayan \& Yi 1995a), the incoming Bondi-Hoyle flow never reaches
$r_{\rm circ}$.  Instead it shocks with the ADAF at a distance on the
order of the outer radius $R_{\rm out}$ (recall that $H\sim R$ in an
ADAF).  In our models we have taken $R_{\rm out}=10^5R_s$, but as we
argued in \S3 the outer radius may be even larger.  The luminosity
associated with the impact at such large radii is quite low.  Once the
stream has been incorporated into the ADAF at the outer radius, the
accretion switches to an advection-dominated form and the gas does not
radiate very much as it moves in.

Several recent papers have made the point that the successful
application of the ADAF model to any observed system is direct
evidence that the accreting object is a black hole (Narayan,
McClintock \& Yi 1996; Narayan, Yi \& Mahadevan 1996; Narayan, Barret
\& McClintock 1997; Narayan, Garcia \& McClintock 1997).  The argument
is that if accretion is via an ADAF and if the object has an event
horizon, then the advected energy will disappear from sight.  However,
if the central object has a surface, then the surface will be heated
by the hot inflow from the ADAF and the advected thermal energy will
be emitted as thermal radiation.  Thus, for an object with a surface,
the radiative efficiency will be the canonical 10\% even if the
accretion occurs via an ADAF.  Only if the central object is a black
hole can the radiative efficiency be truly low.

\sgr is perhaps one of the best objects for this argument.  This
source provides two strong observational constraints: (1) The mass
accretion rate is estimated to be at least a few$\times10^{-6}
M_\odot\,{\rm yr^{-1}}$, and (2) the bolometric luminosity is no
greater than $10^{37}~\ergs$.  In our opinion the only plausible
resolution of these two conflicting pieces of evidence is to postulate
(1) that the accretion in \sgr occurs via an ADAF and (2) that the
central object is a supermassive black hole.  This ``proof'' of the
black hole nature of \sgr is qualitatively different from usual proofs
which rely on a measurement of the mass.  The usual argument is that
if an object is too massive to be a neutron star it must be a black
hole; it is a proof by elimination.  Our ``proof'' is somewhat more
direct and cuts to the essence of what constitutes a black hole,
namely the presence of an event horizon.  We argue that \sgr has an
enormous luminosity deficit for plausible values of $\dot M$ and the
simplest explanation is that the object has an event horizon.

One of the results to come out of this work is our confirmation that
compressive heating of electrons cannot be ignored and must be
included consistently in computations of ADAF spectra of low
luminosity systems like \sgr (see Fig. 2).  This point was made
recently in an important paper by Nakamura et al. (1997).  In view of
this result it would be useful to revisit other low-$\mdot$ systems to
which the ADAF model has been applied (e.g. soft X-ray transients,
Narayan, Barret \& McClintock 1997; Hameury et al. 1997; NGC 4258,
Lasota et al. 1996; low-luminosity nuclei of giant ellipticals, Fabian
\& Rees 1995; Mahadevan 1997; Reynolds et al. 1996) and redo the
analysis with the inclusion of compressive heating.  Interestingly,
Esin et al. (1997) have shown that compressive heating has a much
weaker effect on ADAFs at higher values of $\mdot$.  The reason is
that with increasing $\mdot$ the other terms in the electron energy
equation (1), notably the Coulomb collision term $Q^{\rm ie}$ and the
radiative cooling term $Q^{\rm rad}$, become more important, and
compressive heating no longer dominates.

In the model presented in this paper the emission in the infrared and
the radio (above about 86 GHz) comes from fairly small radii
$\sim10R_s$.  Since the gas in the ADAF is nearly in free-fall, the
characteristic time scales of the flow are quite short.  We may thus
expect rapid variability in \sgr at these wavelengths.  The shortest
likely time scale is the dynamical time, $t_{\rm
dyn}\sim(GM/R^3)^{-1/2}$, which gives $t_{\rm dyn}\sim5000$ s at
$R=10R_s$.  The viscous time is longer than this by a factor of
$1/\alpha \sim3$.  The longer wavelength radio emission is from larger
radii; for instance, the emission below 1 GHz comes from $R\gsim
10^3R_s$.  Variability at these wavelengths will be correspondingly
slower: $t_{\rm dyn}(10^3R_s)\sim5\times10^6$ s.  The bremsstrahlung
emission in soft X-rays is from a broad range of radii, but is
dominated by large radii.  Therefore, the soft X-ray flux should show
much slower variations (timescale $\sim$ 1 year) than the radio,
millimeter or infrared fluxes.  However, the hard X-ray ($\sim100$
keV) flux at the broad peak of the $\nu L_{\nu}$ spectrum will trace
the higher temperature ($\sim10^9-10^{10}$ K) inner regions and could
vary on time scales of days.

By assumption, the electrons in our model are fully thermal.
Mahadevan \& Quataert (1997) have, however, shown that at low mass
accretion rates similar to that in \sgr, electrons may not be
thermalized at larger radii.  The electron energy distribution at
these radii may then be truncated even more sharply than in a
Maxwellian, which will cause a reduction in the synchrotron emission.
This will act to increase the discrepancy between the model and the
data at low radio frequencies.  This is an area for further work.

Note, however, that a power-law distribution of electrons extending
over any reasonable range of energies is constrained by the
observations.  For instance, since optically thin nonthermal
synchrotron emission usually has a spectral form
$L_\nu\sim\nu^{-0.7}$, this means that in a $\nu L_\nu$ plot the
optically thin emission would continue to rise as $\nu^{+0.3}$.  Such
a rise beyond $\sim 10^{12} - 10^{13}$Hz is ruled out by the observed
upper limits in the infrared.

In this connection, we note that Falcke (1996), Duschl \& Lesch (1994)
and Beckert \& Duschl (1997) have proposed nonthermal models for \sgr
in which most of the electrons have Lorentz factors of around a few
hundred.  These authors require a mono-energetic electron distribution
in order to reproduce the sharp cutoff observed in the
sub--millimeter/FIR band.  Falcke assumes a mono-energetic
distribution in the ``nozzle'' of his jet-disk model, while Duschl \&
Lesch assume a homogeneous sphere of mono-energetic electrons.  Such
mono-energetic models are not ruled out by the argument of the
previous paragraph and indeed both groups are able to fit the radio
spectrum of \sgr reasonably well, although the sub-mm, IR and X-ray
spectrum are not explained.  We, however, find the idea of a
nonthermal but mono-energetic distribution somewhat artificial.  Our
ADAF model fits the data in a more natural way by making use of a
thermal distribution of electrons.  Further, because of thermal
synchrotron self-absorption, the cutoff of the spectrum in the
sub-millimeter band is very sharp without requiring any fine--tuning.
Melia's (1992, 1994) model also makes use of self-absorbed thermal
synchrotron emission and has similar properties, but there is an error
in his calculation of the synchrotron emissivity (cf. Mahadevan,
Narayan \& Yi 1996).

Further high spatial resolution observations of \sgr are highly
desirable.  The ADAF model can be tested and constrained with better
observations of the spectrum and variability, especially in the
millimeter, sub--millimeter, infrared and X-ray bands.  Better
$\gamma$-ray observations might also resolve the issue of whether the
source detected by EGRET is Sgr A$^*$.

After this paper was submitted we saw a closely related preprint by
Manmoto, Mineshige \& Kusunose (1997) entitled ``Spectrum of optically
thin advection dominated accretion flow around a black hole:
application to Sgr A$^*$.''  There are many points of similarity
between their paper and ours and fairly close agreement in the
results.  In particular, Manmoto et al. emphasize the importance of
the advection term in the heating of electrons.  The main difference
between the papers is in the choice of equation of state of the
electrons.  We allow for the effect of magnetic fields in calculating
the adiabatic index of the electrons and we also include relativistic
corrections (see Appendix A) whereas Manmoto et al. employ a simpler
prescription.  Also, we make use of a more comprehensive and updated
set of spectral data for the model comparisons.

\acknowledgments This work was supported in part by NSF grant AST
9423209 and NASA grant NAG 5-2837.  We thank Shoji Kato for sending us
preprints of Nakamura et al. (1997) and Manmoto et al. (1997) prior to
publication, and T. Krichbaum, J. Marcaide, S. Stolovy and E. Serabyn
for comments on the spectral data.  RM thanks Jun-Hui Zhao and Mark
Reid for useful discussions on the radio observations of the Galactic
Center.

\vfill\eject
\begin{appendix}
\section{Energy Advection by Electrons}

The advection term in equation (1) can be written per unit volume as
\be
Q^{\rm e,adv}=\rho T_e v \frac{d s_e}{d R},
\ee
where $s_e$ is the entropy of the electrons per unit total gas mass.
This term was ignored in much of the earlier work, on the
assumption that the temperature of the electrons is determined
primarily by a balance between 
the Coulomb transfer term $Q^{\rm ie}$ and the radiative
cooling term $Q^{\rm rad}$ in equation (1).
However, Nakamura et al. (1997) have shown that
energy advection by electrons can be important 
under some circumstances and may even play a dominant role in determining
the electron temperature.  This is the case especially when the
mass accretion rate is low, as in Sgr A$^*$.

We begin with the relation
\be
T_e d s_e = d u_e + P_e d \left(\frac{1}{\rho}\right),
\ee
where $u_e$ is the internal energy of the electrons per unit mass and 
$P_e$ is the electron pressure.  We consider a mixture of gas and
magnetic fields.  If gas pressure contributes a constant fraction
$\beta$ to the total pressure $P_{tot}$, then
\be
P_{tot} = \frac{\rho k T_i}{\mu_i m_u} + \frac{\rho k T_e}{\mu_e m_u} +
\frac{B^2}{24 \pi} = \frac{\rho k T_i}{\beta \mu_i m_u} + 
\frac{\rho k T_e}{\beta \mu_e m_u}.
\ee
It seems natural to denote the two terms 
on the right as the effective ion and 
electron pressure, each including an appropriate fraction of
the magnetic pressure.  Therefore, we have
$P_e = \rho k T_e/(\beta \mu_e m_u)$.

The internal energy of the gas is a sum of the ion, electron and 
magnetic field internal energies:
\be
\label{inten}
u = \frac{3}{2} \frac{k T_i}{\mu_i m_u} + a(T_e) \frac{k T_e}{\mu_e m_u} + 
\frac{B^2}{8 \pi \rho} = \frac{6-3\beta}{2 \beta} \frac{k T_i}{\mu_i m_u} +
\left[\frac{3 (1-\beta)}{\beta} + a(T_e)\right] \frac{k T_e}{\mu_e m_u},
\ee
where the coefficient $a(T_e)$ varies from $3/2$ in the case of a
non-relativistic electron gas, to $3$ for fully relativistic electrons.  
The general expression for $a$ as a function of the dimensionless electron
temperature $\theta_e = k T_e/m_e c^2$ was derived by 
Chandrasekhar (1939, Chapter X, eq.[236]):
\be
a(\theta_e) = \frac{1}{\theta_e} \left(\frac{3 K_3 (1/\theta_e) + 
K_1 (1/\theta_e)}{4 K_2 (1/\theta_e)} - 1\right).
\ee
Note that the ions never become relativistic in these flows,
so that the corresponding coefficient for the ions is
always $\sim 3/2$.

As we have done for the pressure, the right hand side of 
Eq. (\ref{inten}) may be naturally divided into two terms, the ion 
and electron internal energies:
\be
u_i = \frac{6-3\beta}{2 \beta} \frac{k T_i}{\mu_i m_u},\ \ \ {\rm and}\ \ \ 
u_e= \left[\frac{3 (1-\beta)}{\beta}+ a(T_e)\right] \frac{k T_e}{\mu_e m_u}.
\ee
In this interpretation, $u_i$ and $u_e$ are again ``effective'' quantities, 
which include contributions from the particles as well as the associated 
magnetic field.  Note that the contribution 
of the magnetic field to the internal energy of each particle 
species is proportional to the contribution of these particles to the total 
pressure, a natural choice in our model where the ratio of the magnetic
to gas pressure is fixed.
Having defined $P_e$ and $u_e$, we can now write the energy
advection term for the electrons as
\be
Q^{\rm e,adv}=
\frac{\rho v k}{\mu_e m_u} \left[\frac{3 (1-\beta)}{\beta} + a(T_e) + 
T_e \frac{d a}{d T_e}\right] \frac{d T_e}{d R} - \frac{v k T_e}{\beta
\mu_e m_u} \frac{d \rho}{d R}.
\ee

>From the pressure and internal energy of the electrons, we calculate the
effective adiabatic index of the electrons via the relation
$\gamma_e-1=P_e/u_e$.  This gives 
\be 
\gamma_e =\frac{4-3\beta+a\beta}{3-3\beta+a\beta}.  
\ee 
If the
particles are relativistic, then $a=3$ and $\gamma_e=4/3$ regardless
of $\beta$.  This is because both the particles and the tangled field
behave like radiation.  If the particles are non-relativistic, however,  
$a=3/2$ and in this case 
\be
\gamma_e=\frac{8-3\beta}{6-3\beta},
\ee
as shown by Esin (1997).  We see that $\gamma_e$
varies from 4/3 when $\beta=0$ to 5/3 when $\beta=1$.  Thus,
the adiabatic index depends fairly sensitively on $\beta$ so long
as the electrons are non-relativistic.
If the electrons in the accreting gas behave adiabatically, then
their temperature varies as $T_e\propto\rho^{\gamma_e-1}$ (technically
there is also the term proportional to $da/dT_e$ in eq A7 which becomes
important when the electrons are quasi-relativistic, but we
neglect this for simplicity).  Thus, for a given amount of compression,
the electrons are hotter when $\beta$ is large and cooler when
$\beta$ is small.  This explains the trend seen in Figs. 4b and 4c.

\end{appendix}

\newpage

\noindent{\bf Figure Captions}

\bigskip
\noindent
Figure~1.  The open and filled circles represent various flux
measurements and upper limits of Sgr A$^*$.  We consider the filled
circles to be more important as model constraints.  The box at 1 keV
represents the uncertainty in the X-ray flux.  The solid line is our
baseline ADAF model with the following parameters:
$M=2.5\times10^6M_\odot$, $\alpha=0.3$, $\beta=0.5$, $\delta=0.001$.
The Eddington-scaled mass accretion rate $\mdot$ has been adjusted to
fit the X-ray flux, giving $\mdot=1.3\times10^{-4}$.  The peak at the
left is due to synchrotron radiation, the next peak is due to Compton
scattering, the peak between 10--100 keV is due to bremsstrahlung and
the peak above 100 MeV is due to pion production.  The long--dashed
line is a model in which the pion peak has been artificially raised by
about an order of magnitude to fit the data.  The dotted line is the
spectrum corresponding to a standard thin accretion disk with
$\mdot=10^{-4}$ while the short--dashed line is a thin disk with
$\mdot=10^{-9}$.  Neither of these models is satisfactory.

\noindent 
Figure~2.  (a) Spectra corresponding to five models, showing the effect
of various approximations; see the text for details.  (b) Electron
temperature profiles corresponding to the same five models.  The
models represented with long-dashed and solid lines have nearly
identical temperatures.

\noindent
Figure ~3.  (a) The solid line is the baseline model shown in Fig. 1,
with a black hole mass of $M=2.5\times10^6M_\odot$.  The dotted line
corresponds to a model with $M=8.3\times10^5M_\odot$ and the dashed
line corresponds to $M=7.5\times10^6M_\odot$.  (b) The solid line is
the baseline model with a mass accretion rate
$\mdot=1.3\times10^{-4}$.  The dotted, short-dashed, long-dashed and
dash-dotted lines corresponds to models with
$\mdot=6.4\times10^{-5}, ~9.0\times10^{-5}, ~1.8\times10^{-4},
~\mdot=2.5\times10^{-4}$, respectively.

\noindent
Figure ~4.  (a) The solid line is the baseline model shown in Fig. 1,
with $\alpha=0.3$.  The dotted, short-dashed and long-dashed lines
correspond to models with $\alpha=0.1, ~0.2, ~0.4$ respectively.  (b)
The solid line is the baseline model shown in Fig. 1, with
$\beta=0.5$.  The dotted, short-dashed, long-dashed lines and
dot-dashed lines correspond to models with $\beta=0.3, ~0.4, ~0.6,
~0.7$ respectively.  (c) Electron temperature profiles corresponding
to the models shown in (b).  (d) The solid line is the baseline model
shown in Fig. 1, with $\delta=0.001$.  The dotted, short-dashed,
long-dashed and dot-dashed lines correspond to models with $\delta=0,
~0.003162, ~0.01, ~0.03162$ respectively.

\noindent
Figure ~5.  (a) The solid line is the baseline model shown in Fig. 1.
The dotted line corresponds to a model in which the temperature of the
electrons has been fixed at $T_e=2\times10^9$ K over the radius range
$r=20-1000$.  This model fits the low frequency radio data better.
(b) Electron temperature profiles corresponding to the two models
shown in (a).

\newpage
\begin{table}
\caption[tab 1]
{Sgr A$^*$: Radio and NIR Observations (Distance = 8.5 kpc)}
\vskip-3.0cm
\begin{center}
\begin{tabular}{lccccl}\hline 
& & & & &  \\
\rb{$\nu$} & \rb{$\lambda$}& \rb{$\theta$} & \rb{$S_{\nu}$} & \rb{$\nu L_{\nu}$} & \rb
{Ref.} \\
\rb{Hz} &\rb{$\mu$m}& \rb{$ \prime \prime$} &\rb{Jy}& \rb{ergs s$^{-1}$} & \\[-.5ex]
\hline \hline
4.08$\times 10^{8}$ & $735294$ & $4.3$ & $\le 0.05$& $\le 1.76 \times 10^{30}$& Davies et al. 76 \\
9.6$\times 10^{8}$ & $312500$ & $10$ & $0.29$& $ 2.41 \times 10^{31}$& Davies et al. 76 \\
9.6$\times 10^{8}$ & $312500$ & $10$ & $0.27$& $ 2.24 \times 10^{31}$& Davies et al. 76 \\
9.6$\times 10^{8}$ & $312500$ & $10$ & $0.30$& $ 2.49 \times 10^{31}$& Davies et al. 76 \\
1.66$\times 10^{9}$ & $180722$ & $2.5$ & 0.56& $8.04\times 10^{31}$& Davies et al. 76 \\
1.5 $\times 10^{9}$ & $200000$ & $\sim 1.2$ & 0.8& 1.04$\times 10^{32}$& Backer 82 \\
1.5 $\times 10^{9}$ & $200000$ & $\sim 1.2$ & 0.3& 3.89$\times 10^{31}$& Zhao et al. 89 \\
2.3 $\times 10^{9}$ & 130000 & $ 0.2$ & 1.1$\pm0.1$& $(2.2\pm0.2) \times 10^{32}$& Marcaide et al. 92 \\
2.7 $\times 10^{9}$ & $110000$ & $\sim 0.65$  & .73& $ 1.7 \times 10^{32}$& Backer 82 \\
2.7 $\times 10^{9}$ & $110000$ & $\sim 0.65$  & .42& $ 9.78 \times 10^{31}$& Brown \& Lo  82\\
5.0 $\times 10^{9}$ & 60000 & $\sim 0.35$ & 1.13& 1.45$\times 10^{32}$& Zhao et al. 89 \\
5.0 $\times 10^{9}$ & 60000 & $\sim$ 0.35 & 0.55& 7.13$\times 10^{31}$& Zhao et al. 89 \\
8.1 $\times 10^{9}$ & 37000 & $\sim 0.22$ & 0.9& 6.23$\times 10^{32}$& Backer 82 \\
8.1 $\times 10^{9}$ & 37000 & $\sim 0.22$ & 0.58& $4.01 \times 10^{32}$& Brown \& Lo 82 \\
8.3 $\times 10^{9}$ & 36000 & $ 0.016$ & 1.2$\pm0.1$& $(8.6\pm0.7) \times 10^{32}$& Marcaide et al. 92 \\
8.4 $\times 10^{9}$ & 35714 & $\sim 0.21$ & 1.07& $ 7.77 \times 10^{32}$& Zhao et al. 92 \\
8.4 $\times 10^{9}$ & 35714 & $\sim 0.21$ & 0.55& $3.99 \times 10^{32}$& Zhao et al. 92 \\
1.5 $\times 10^{10}$ & 20000 & $\sim 0.12$& 1.64& 2.1$\times 10^{33}$& Zhao et al. 92 \\
1.5 $\times 10^{10}$ & 20000 & $\sim 0.12$& .68& 8.7$\times 10^{32}$& Zhao et al. 92 \\
\end{tabular}
\end{center}
\end{table}
\newpage
\addtocounter{table}{-1}
\begin{table}
\caption[]
{ Sgr A$^*$: Radio and NIR Observations (continued).}
\vskip-3.0cm
\begin{center}
\begin{tabular}{lccccl}\hline 
& & & & &  \\
\rb{$\nu$} & \rb{$\lambda$}& \rb{$\theta$} & \rb{$S_{\nu}$} & \rb{$\nu L_{\nu}$} & \rb
{Ref.} \\
\rb{Hz} &\rb{$\mu$m}& \rb{$ \prime \prime$} &\rb{Jy}& \rb{ergs s$^{-1}$} & \\[-.5ex]
\hline \hline
1.5 $\times 10^{10}$ & 20000 & 0.12$\times$0.24 & 1.15+0.01& (1.49+0.06)$\times 10^{33}$& Yusef--Zadeh et al. 90 \\
2.2 $\times 10^{10}$ & 13600 & $\sim 0.08$& 2.1& $3.99\times 10^{33}$& Zhao et al. 92 \\
2.2 $\times 10^{10}$ & 13600 & $\sim 0.08$& 0.8& 1.52$\times 10^{33}$& Zhao et al. 92 \\
2.2$\times 10^{10}$ &13500& 1.8$\times 10^{-3}$& $1.07\pm 0.15$ &$(2.04\pm 0.29)\times 10^{33}$ & Alberdi et al. 93 \\
2.3$\times 10^{10}$ &13000& 1.8$\times 10^{-3}$& $1.2\pm 0.4$ &$(2.4\pm 0.8)\times 10^{33}$ & Marcaide et al. 92 \\
\hspace{-.29cm}$\!\star$ 4.3 $\times 10^{10}$ & 7000 & 0.75$\times 10^{-3}$ & 1.4$\pm 0.1$& $(5.2\pm 0.4)\times 10^{33}$& Krichbaum et al. 94 \\
8.6 $\times 10^{10}$ & 3488 & 0.02 & 1.3& 9.67$\times 10^{33}$& Backer 82 \\
8.6 $\times 10^{10}$ & 3488 & 4$\times8$ & 1.05& 7.8$\times 10^{33}$& Wright et al. 87 \\
\hspace{-.29cm}$\!\star$ 8.6 $\times 10^{10}$ & 3488 & $0.16 \times 10^{-3}$ & 1.4$\pm0.2$&
$(1.04\pm 0.15)\times 10^{34}$& Rogers et al. 94 \\
2.2$\times 10^{11}$ &1350& 1.9$\times 4.3$& $2.4\pm 0.5$ &$(4.57\pm .95)\times 10^{34}$ & Serabyn et al. 92 \\
2.3$\times 10^{11}$ &1300& 11& $2.5$ &$4.97\times 10^{34}$ & Zylka \& Mezger 88 \\
2.3$\times 10^{11}$ &1300& 11& $2.6\pm0.6$ &$(5.2\pm 1.2)\times 10^{34}$ & Zylka et al. 92 \\
3.5$\times 10^{11}$ &870& 8 & $4.8\pm 1.2$ &$(1.45\pm 0.36)\times 10^{35}$ & Zylka et al. 92 \\
3.75$\times 10^{11}$ &800& 13 & $3.5\pm 0.5$ &$(1.13\pm 0.16)\times 10^{35}$ & Zylka et al. 95\\
5.0$\times 10^{11}$ &600& 10 & $4.0\pm 1.2$ &$(5.19\pm 0.52)\times 10^{35}$ & Zylka et al. 95\\
6.7$\times 10^{11}$ &450& 7 & $\le 1.5$ &$ \le 8.69 \times 10^{34}$ & Dent et al. 93\\
6.7$\times 10^{11}$ &450& 8 & $3.0\pm 1.0$ &$(1.74\pm 0.58)\times 10^{35}$ & Zylka et al. 95\\
8.6$\times 10^{11}$ &350& 11 & $7\pm2$ & $(5.2\pm1.5)\times10^{35}$ &
Serabyn et al. 97 \\
8.6$\times 10^{11}$ &350& 11 & $\le 10$ &$\le 7.4\times 10^{35}$ & Mezger 1994\\
8.6$\times 10^{11}$ &350& 30& $18.5\pm 9$ &$(1.38\pm 0.67)\times 10^{36}$ & Zylka et al. 92 \\
1.0$\times 10^{13}$ &30& 8 & $\le 120$ &$\le 1.04 \times  10^{38}$ & Zylka et al. 92 \\
\end{tabular}
\end{center}
\end{table}
\newpage
\addtocounter{table}{-1}
\begin{table}
\caption[]
{ Sgr A$^*$: Radio and NIR Observations (continued).}
\vskip-3.0cm
\begin{center}
\begin{tabular}{lccccl}\hline 
& & & & &  \\
\rb{$\nu$} & \rb{$\lambda$}& \rb{$\theta$} & \rb{$S_{\nu}$} & \rb{$\nu L_{\nu}$} & \rb
{Ref.} \\
\rb{Hz} &\rb{$\mu$m}& \rb{$ \prime \prime$} &\rb{Jy}& \rb{ergs s$^{-1}$} & \\[-.5ex]
\hline \hline
1.0$\times 10^{13}$ &30& 4 & $\le 20$ &$\le 1.7 \times  10^{37}$ &  Telesco et al. 96 \\
\hspace{-.29cm}$\!\star$ 1.5$\times 10^{13}$ &20& 1.6 & $\le 1$ &$ \le 1.3 \times 10^{36}$ & Gezari et al. 94\\
1.56$\times 10^{13}$ &19.2& 4 & $\le 1.4$ &$ \le 1.9 \times 10^{36}$ & Telesco et al. 96\\
\hspace{-.29cm}$\!\star$ 1.7$\times 10^{13}$ &18& 2.3 $\times 1.3$ & $\le 0.3$ &$\le 4.4\times 10^{35}$ & in Zylka et al. 92 \\
 2.3--3.6$\times 10^{13}$ &13--8& 2.3 $\times 1.3$ & $\le 0.1$ &$\le 2.6\times 10^{35}$ & in Zylka et al. 92 \\
\hspace{-.29cm}$\!\star$ 3.5$\times 10^{13}$ &8.7& 0.5 & $\le 0.1$ & $\le 3.0
\times10^{35}$ & Stolovy et al. 96 \\
\hspace{-.29cm}$\!\star$ 1.4$\times 10^{14}$ &2.2& 0.15 &$\le 9\times 10^{-3}$ & $\le 1.1\times 10^{35}$ & Menten et al. 97 \\
\end{tabular}
\end{center}
\end{table}
\begin{table}
\caption[]
{ Sgr A$^*$: X--Ray and $\gamma$--Ray Observations (Distance = 8.5 kpc).}
\begin{center}
\begin{tabular}{lccccl}\hline 
& & & & &  \\
\rb{Energy} & \rb{Telescope/}& \rb{$\theta$}& 
\rb{$L_{\rm EB}$ \footnotemark[1]}
& \rb{$\nu L_{\nu}$} 
& \rb{Ref.} \\
\rb{Band (EB)}&\rb{Instrument} & &\rb{erg s$^{-1}$} & \rb{erg 
s$^{-1}$} &  \\[-.5ex]
\hline \hline
0.8 - 2.5 keV& ROSAT& $\sim 20 \arcsec$ & $ 1.55  \times 10^{34}$
& 1.6 $\times 10^{34}$ & Predehl \& Tr\"umper 94 \footnotemark[2] \\
2 - 10 keV& ASCA & $\sim 1\arcmin$ &$\le 6.4 \times 10^{35}$ & 
$\le 4.8 \times 10^{35}$ & Koyama et al. 96 \\ 
35 - 75 keV &SIGMA &$\sim 15 \arcmin$& $\le 3.5\times 10^{35}$ & 
$\le 4.8 \times 10^{35}$ &  Goldwrum et al. 94 \\
75 - 150 keV &SIGMA & $\sim 15 \arcmin$& $ \le 2.4 \times 10^{35}$&
$\le 3.6 \times 10^{35}$&  Goldwrum et al. 94 \\
30 - 50 MeV& EGRET & $\sim 1\degree$& & $\le 1.8 \times 10^{36}$ &Merck et al. 96\\
50 - 70 MeV& EGRET & $\sim 1\degree$& & $\le 1.1 \times 10^{36}$ &Merck et al. 96\\
70 - 100 MeV& EGRET & $\sim 1\degree$& & $\le 8.3 \times 10^{35}$ &Merck et al. 96\\
100 - 150 MeV& EGRET & $\sim 1\degree$& & $\le 6.2 \times 10^{35}$ &Merck et al. 96\\
150 - 300 MeV& EGRET & $\sim 1\degree$& & $(4.9^{+2.0}_{-2.1}) \times 10^{35}$&Merck et al. 96\\
300 - 500 MeV& EGRET & $\sim 1\degree$& & $(1.2^{+0.24}_{-0.26}) \times 10^{36}$ &Merck et al. 96\\
500 - 1000 MeV& EGRET & $\sim 1\degree$& & $(1.4^{+0.28}_{-0.28}) \times 10^{36}$ &Merck et al. 96\\
1 - 2 GeV& EGRET & $\sim 1\degree$& & $(1.7^{+0.28}_{0.28}) \times 10^{36}$ &Merck et al. 96\\
2 - 4 GeV& EGRET & $\sim 1\degree$& & $(2.2^{+0.82}_{-0.62}) \times 10^{36}$ &Merck et al. 96\\
4 - 10 GeV& EGRET & $\sim 1\degree$& & $(8.3^{+4.2}_{-4.7}) \times 10^{35}$ &Merck et al. 96\\
& &&&& \\
\multicolumn{6}{l}{
 \footnotesize{$^{1}L_{\rm EB}$ is the total luminosity integrated over the band.}} \\
\multicolumn{6}{l}{\footnotesize{$^2$ This flux is obtained using 
$N_H = 6\times 10^{22} $cm$^{-2}$ as opposed to the much higher }} \\
\multicolumn{6}{l}{\footnotesize{
column density used by Predehl \& Trumper (1994). This is discussed 
in \S 2.}}
\end{tabular}
\end{center}
\end{table}
\newpage
\pagestyle{empty}
\begin{figure}
\epsffile{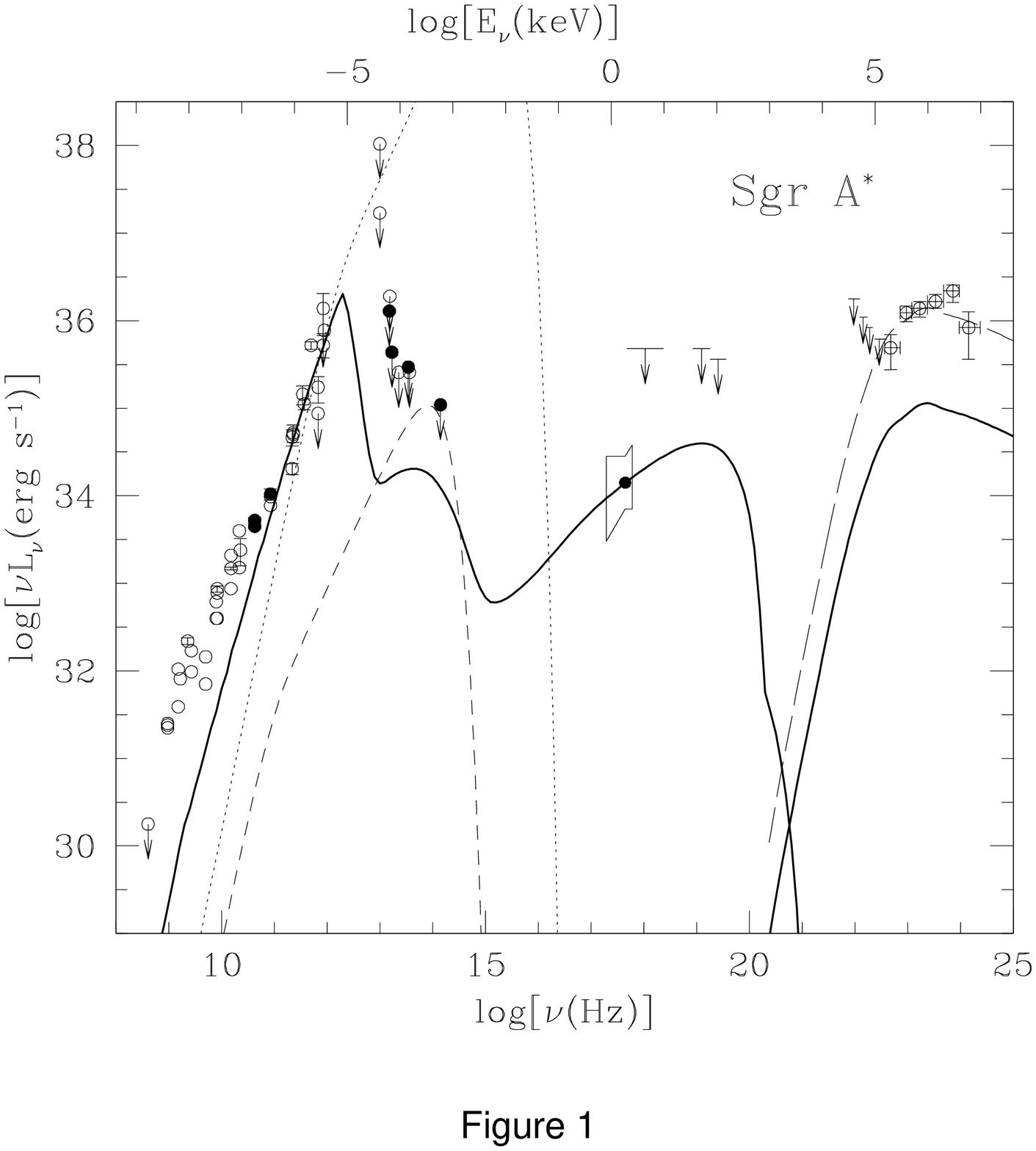}
\end{figure}
\begin{figure}
\epsffile{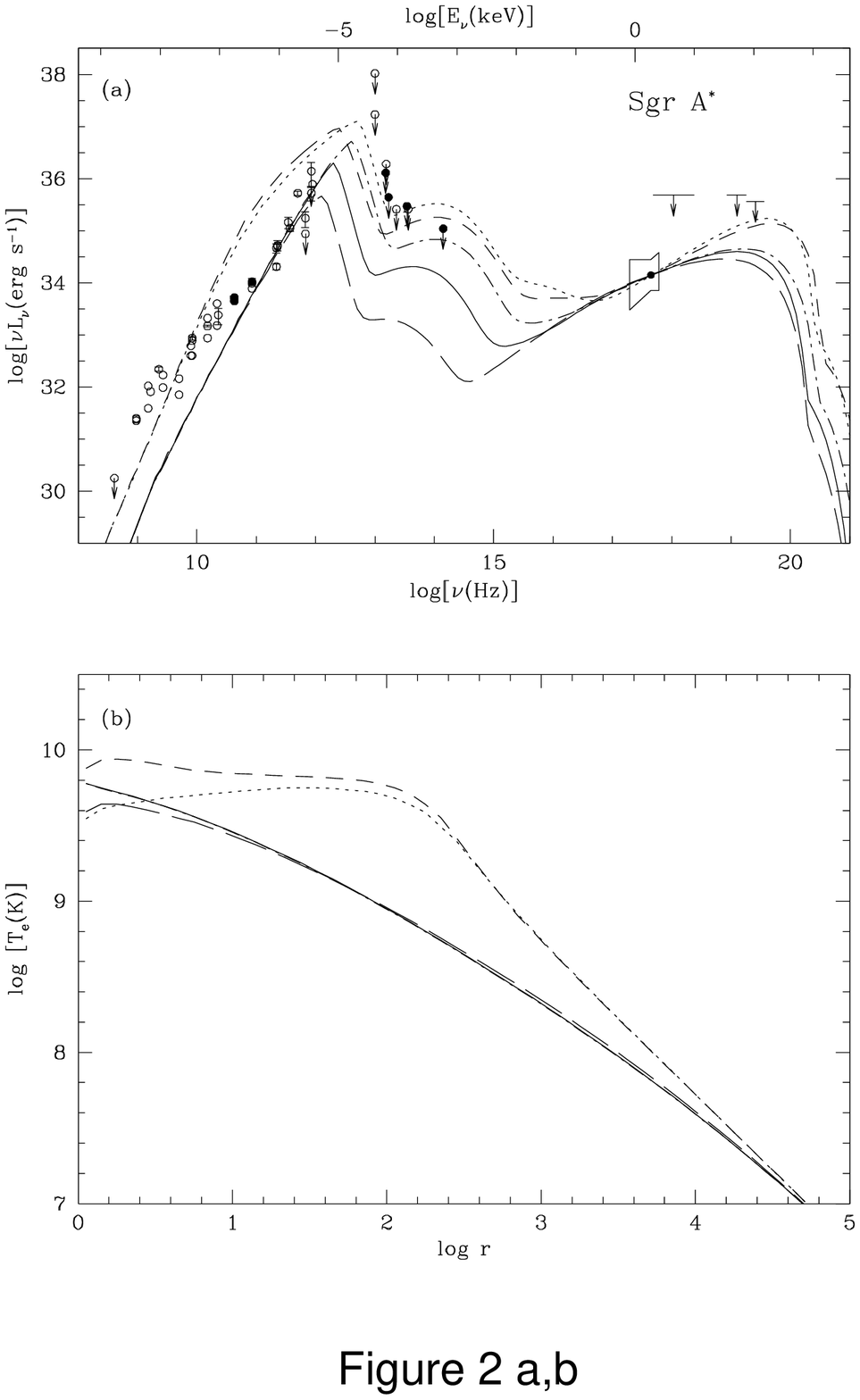}
\end{figure}
\begin{figure}
\epsffile{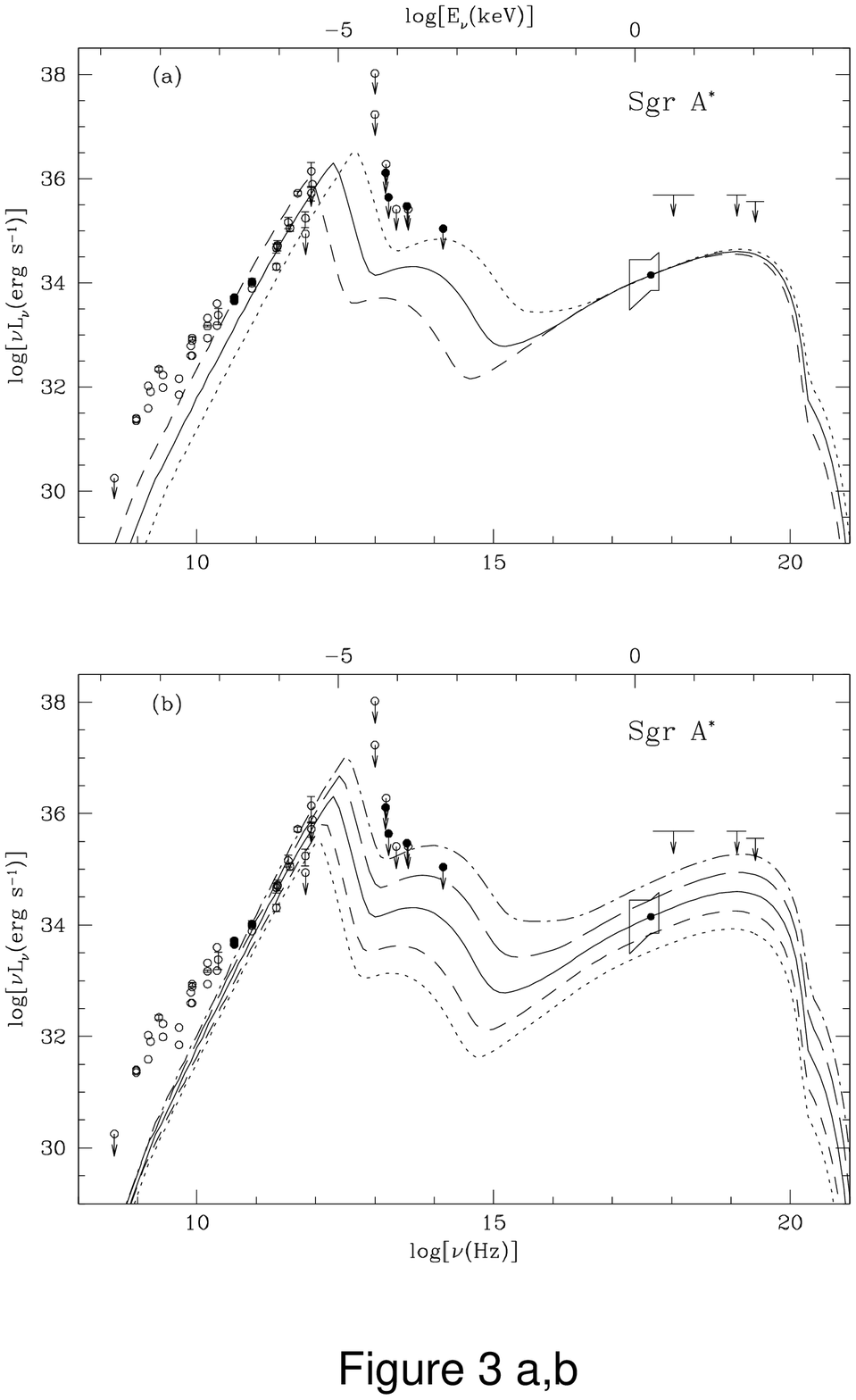}
\end{figure}
\begin{figure}
\epsffile{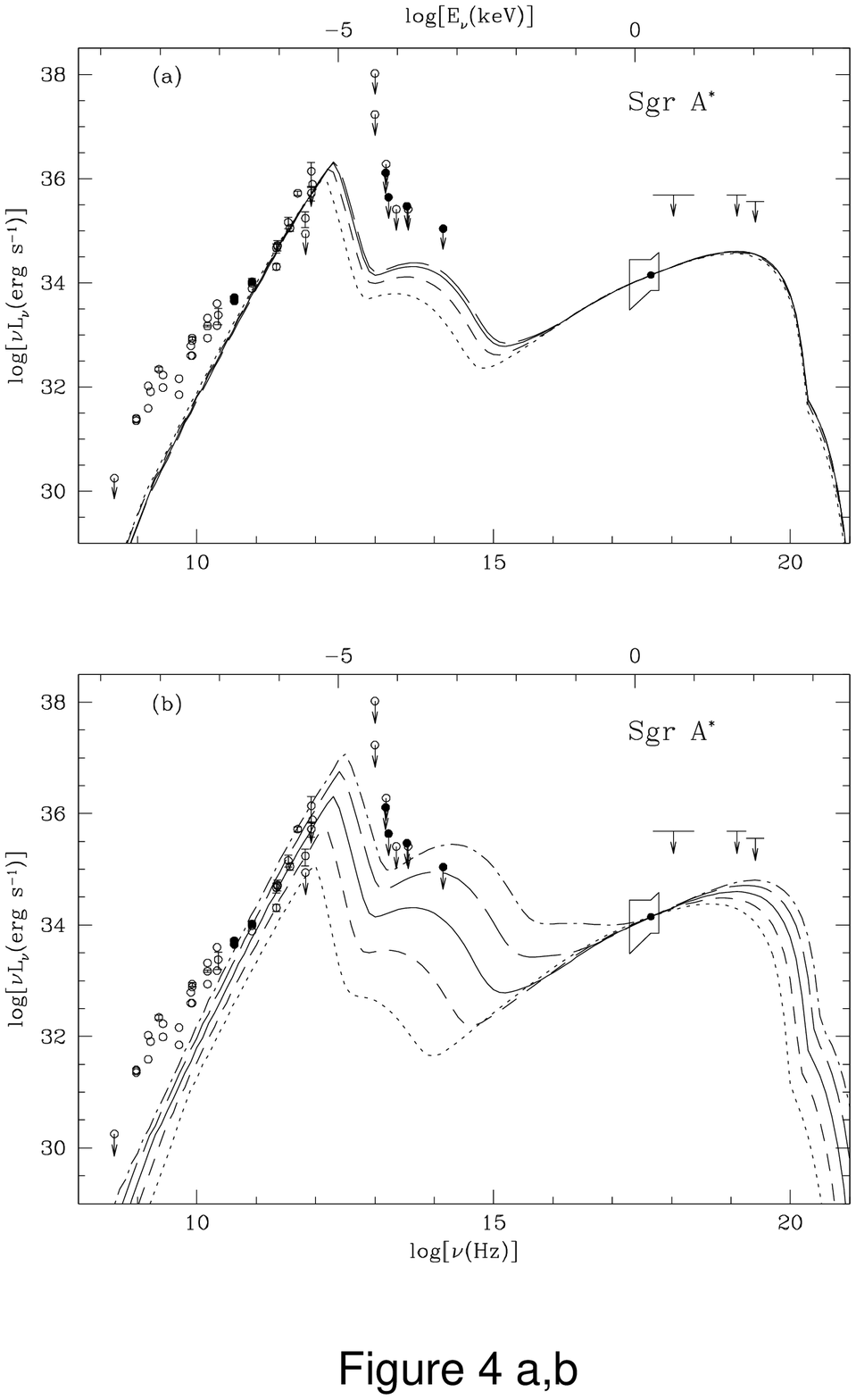}
\end{figure}
\begin{figure}
\epsffile{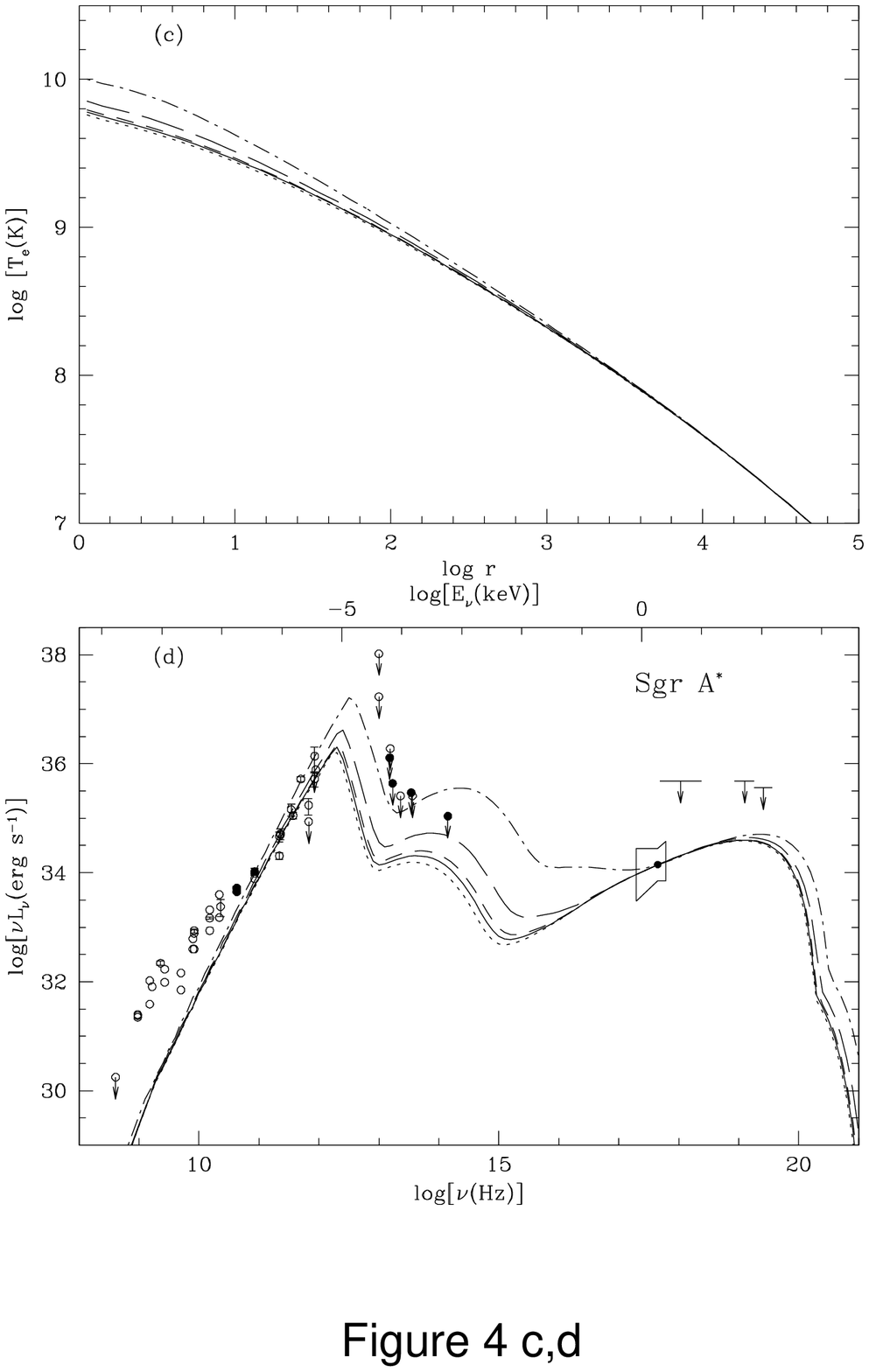}
\end{figure}
\begin{figure}
\epsffile{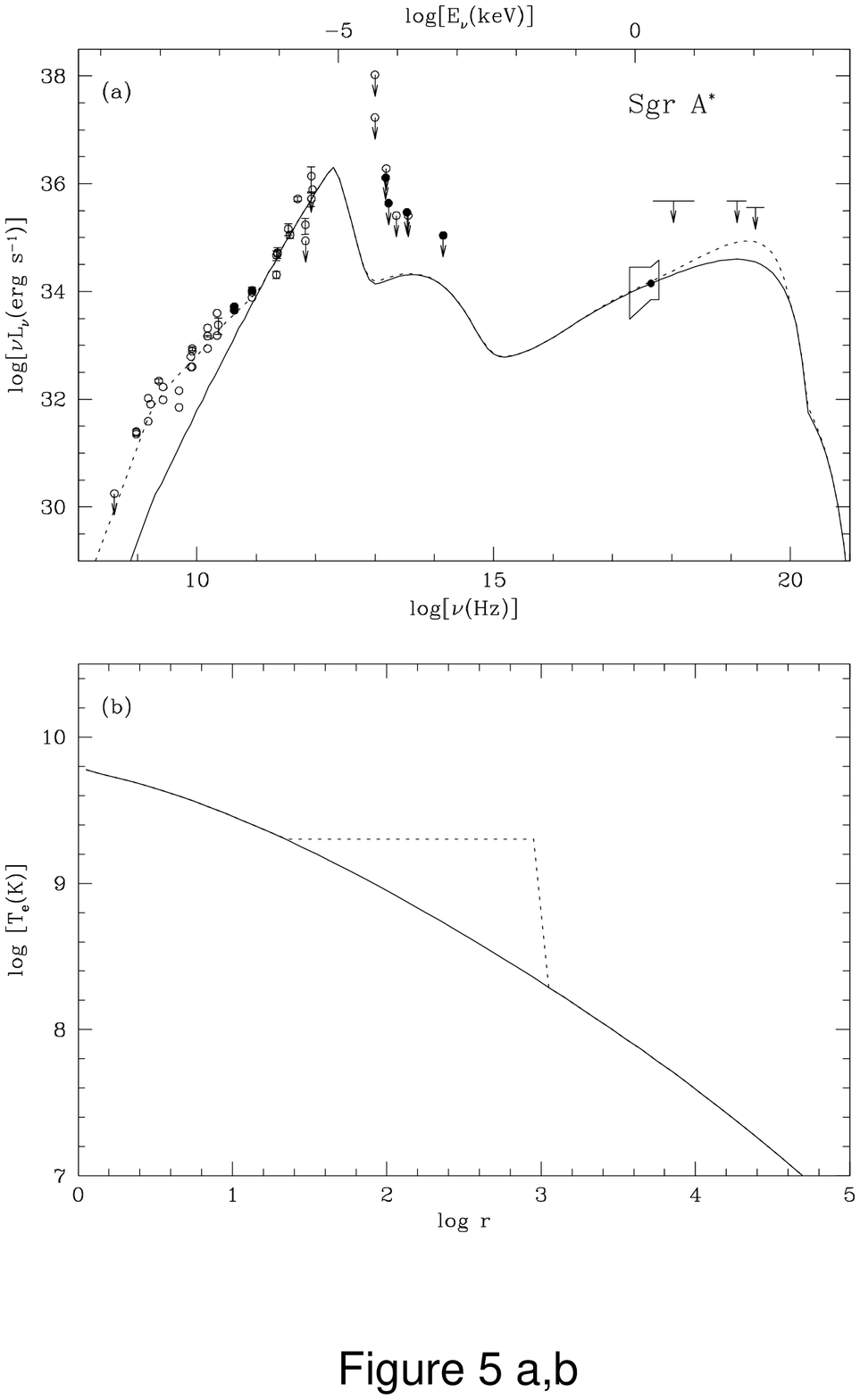}
\end{figure}
\end{document}